\documentclass[prx,reprint]{revtex4-1}
\usepackage{graphicx}
\usepackage{tocvsec2}
\usepackage{amsmath}
\usepackage{amssymb}
\def\myr{\mathrm}
\newcommand{\abs}[1]{\left|#1\right|}
\newcommand{\Be}{$^9$Be$^+$~}
\newcommand{\Hg}{$^{198}$Hg$^+$~}
\newcommand{\un}[1]{\mathrm{#1}}

\newcommand{\capfont}[1]{{(\bf #1)}}

\newcommand{\Ud}{U_{\myr{depth}}}
\newcommand{\Eref}[1]{Eq.~(\ref{#1})}
\newcommand{\Fref}[1]{Fig.~\ref{#1}}
\newcommand{\Frefs}[1]{Figs.~\ref{#1}}
\newcommand{\angstrom}{{\textup{\AA}}}
\newcommand{\Sref}[1]{Sec.~\ref{#1}}
\newcommand{\Aref}[1]{appendix~\ref{#1}}
\newcommand{\eref}[1]{(\ref{#1})}
\newcommand{\Tref}[1]{table~\ref{#1}}
\begin{document}

\title[]{Hybrid quantum systems with trapped charged particles}

\author{Shlomi Kotler}
\email{shlomi.kotler@nist.gov}
\author{Raymond W. Simmonds}\author{Dietrich Leibfried}\author{David J. Wineland}
\affiliation{National Institute of Standards of Technology, 325 Broadway St., Boulder, CO 80305.}

\date{\today}
\begin{abstract}
	Trapped charged particles have been at the forefront of quantum information processing (QIP) for a few decades now, with deterministic two-qubit logic gates reaching record fidelities of $99.9\%$ and single qubit operations of much higher fidelity. In a hybrid system involving trapped charges, quantum degrees of freedom of macroscopic objects such as bulk acoustic resonators, superconducting circuits or nano-mechanical membranes, couple to the trapped charges and ideally inherit the coherent properties of the charges. The hybrid system therefore implements a ``quantum transducer", where the quantum reality (i.e. superpositions and entanglement) of small objects is extended to include the larger object. Although a hybrid quantum system with trapped charges could be valuable both for fundamental research and for QIP application, no such system exists today. Here we study theoretically the possibilities of coupling the quantum mechanical motion of a trapped charged particle (e.g. ion or electron) to quantum degrees of freedom of superconducting devices, nano-mechanical resonators and quartz bulk acoustic wave resonators. For each case, we estimate the coupling rate between the charged particle and its macroscopic counterpart and compare it to the decoherence rate, i.e. the rate at which quantum superposition decays. A hybrid system can only be considered quantum if the coupling rate significantly exceeds all decoherence rates. Our approach is to examine specific examples, using parameters that are experimentally attainable in the foreseeable future. We conclude that those hybrid quantum system considered involving an atomic ion are unfavorable, compared to using an electron, since the coupling rates between the charged particle and its counterpart are slower than the expected decoherence rates. A system based on trapped electrons, on the other hand, might have coupling rates which significantly exceed decoherence rates. Moreover it might have appealing properties such as fast entangling gates, long coherence and flexible electron interconnectivity topology. Realizing such a system, however, is technologically challenging, since it requires accommodating both trapping technology and superconducting circuitry in a compatible manner. We review some of the challenges involved, such as the required trap parameters, electron sources, electrical circuitry and cooling schemes in order to promote further investigations towards the realization of such a hybrid system. 
\end{abstract}

\pacs{37.10.Ty,77.65.Fs,85.25.-j}

\maketitle
\tableofcontents

\section{Introduction}
\begin{figure*}
	\includegraphics[scale=0.8]{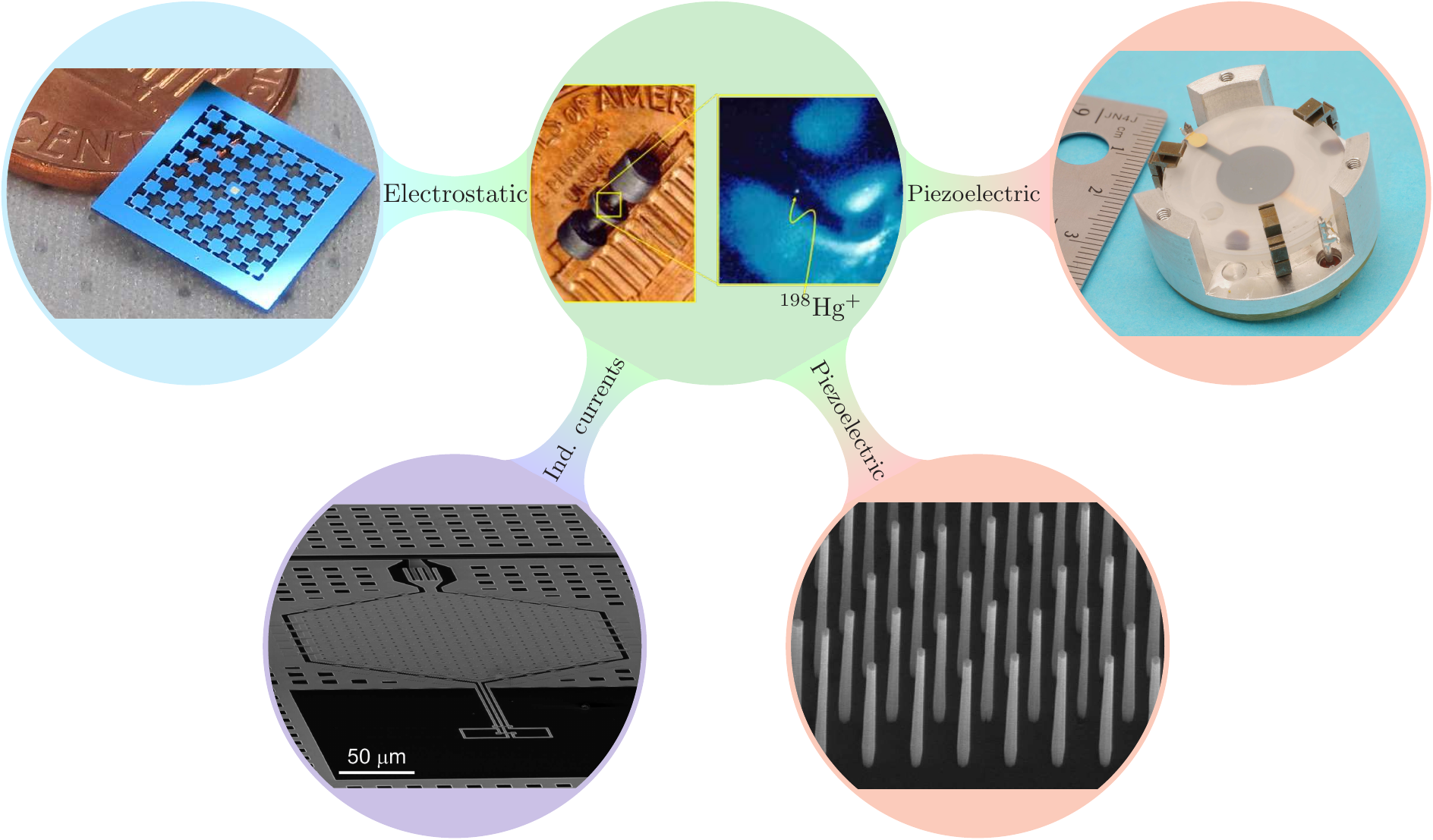}
	\caption{Examples of different platform candidates for a hybrid architecture considered in this paper. Clockwise from the top: \Hg ion trap,  quartz bulk acoustic wave resonator, gallium nitride nano-beams, superconducting LC circuit, nano-mechanical silicon nitride (SiN) membrane. The ion (green shading) is coupled via piezoelectricity (red shading), induced image currents (purple shading) or electrostatics (blue shading). Ion trap photo courtsey of J. Bergquist, NIST, Boulder, CO 80305, USA. Gallium nitride nano-beams photos courtesy of K. Bertness, NIST, Boulder, CO 80305, USA. Quartz resonator device courtesy of S. Galliou, FEMTO-ST institute, 25000 Besan\c{c}on, France. SiN membrane photo courtesy of K. Cicak, NIST, Boulder, CO 80305, USA.} \label{fig:mechplatforms}
\end{figure*}
Trapping of charged particles~\cite{Paul1990,Dehmelt:1990ko} has enabled long interrogation times of their external and internal states, enabling precision metrology, such as in atomic clocks. Applying these tools to atomic ions, paired with the ability of laser-enabled state manipulation, can also turn ions into a quantum information processing (QIP) platform~\cite{Blatt:2008gj,Hanneke:2010iu,Schindler:2013hu,Monroe:2013hq,Roos:2014hm}. Ions have demonstrated record fidelities for initialization, readout, individual spin manipulation~\cite{Harty:2014cm} and entanglement~\cite{Ballance:2016hy,Gaebler:2016ge}.

Other quantum-coherent systems might therefore benefit, by coupling to trapped ions, potentially inheriting aspects of their high controllability and coherence. For example, as described below, one might be able to use a single \Be ion coupled to a $\sim 10~\un{mg}$ quartz resonator to cool the latter close to its ground state. By placing the ion in a superposition state of motion and transferring it to a macroscopic resonator one could explore bounds on quantum mechanics for massive objects. The ion therefore could provide a ``quantum transducer" that enables the manipulation of a much larger object in a coherent way at the single phonon level. For the purpose of QIP, ions might be used as excellent memory units, e.g. for superconducting devices, as long as quantum information can be exchanged between the two systems on time scales that are sufficiently short compared to the decoherence time of the superconducting circuit. The internal degrees of freedom of an ion can remain coherent for tens to hundreds of seconds~\cite{Bollinger:1991cv,Fisk:1995hv,Langer:2005hn,Harty:2014cm}, significantly exceeding the lifetime of coherent excitation in current superconducting devices, typically limited to below $100\ \mu\un{s}$ (e.g. see~\cite{Geerlings:2012ju}), setting the time-scale for useful quantum exchange.

The resonant interaction of ions with radio frequency electrical resonators was studied in~\cite{Heinzen:1990zz}. Complementary parametric interaction schemes for the non-resonant case were recently studied in~\cite{Wineland:1973gfa,Kielpinski:2012hw,Daniilidis:2013kw,Kafri:2015uf,DeMotte:2015wy}. Other suggestions include interfacing nano-mechanical resonators~\cite{Wineland:335673,Tian:2004ko,Hensinger:2005kc,Hunger:2011eo,Daniilidis:2013jz}, electrical wires~\cite{Daniilidis:2009ga} and superconducting qubits~\cite{Daniilidis:2013jz}. These reports analyzed the basic physics involved in each of the different coupling mechanisms as well as the prospects of using such hybrid systems.

Here we focus on a few specific examples of hybrid systems rather than presenting a general treatment. For these examples we take into account available materials, achievable quality factors and practical limitations. Nevertheless, our analysis is based on a unified framework (\Sref{sec:butterworth}), that allows for direct comparison of relevant figures of merit associated with the different systems.  We hope these examples are representative of the different opportunities available and illuminate some of the issues of hybrid QIP with charged particles.

A charged particle moving in a harmonic trap gives rise to an oscillating electric dipole. This dipole in turn can couple to nearby charged objects~\cite{Wineland:335673,Brown:2011kj,Harlander:2011hc}, generate image currents in a nearby conductor~\cite{Heinzen:1990zz}, polarize a dielectric material, or induce motion in a piezo-electric crystal. If the coupled system also has a harmonic mode resonant with the ion motion, energy exchange will occur between the ion harmonic motion and the coupled system. 

The analysis that follows below is guided by the realization that coupling two quantum systems is a double edged sword. Ideally one would like to benefit from the useful properties of both systems involved. In reality, the hybrid system often inherits the disadvantages of both constituents. Therefore, to retain any useful quantum characteristics, we require that the coupling rate between the two systems exceeds the fastest relevant decoherence rate in both systems. Additionally, we focus on specific architectures where the two technologies involved could be compatible and not preclude either of the coupled systems from being close to a pure quantum state.

Although we cannot completely rule out all mechanisms considered here that involve an atomic ion, the analysis emphasizes how challenging it would be to incorporate one into a hybrid system at the quantum level. The coupling rates we calculate, based on experimentally attainable parameters, are either well below the decoherence rates or marginally close to them. This conclusion changes when considering coupling a charged particle to a superconducting resonator, assuming an electron rather than an ion. This follows from the fact that for a particle of mass $m$ the coupling rate is proportional to $m^{-1/2}$ (see~\Sref{sec:supercondresonator}), rendering coupling rates on the order of $\sim 0.1-1~\un{MHz}$ where we expect to exceed decoherence rates. 

The shift from using an atomic ion to using an electron has significant practical implications as detailed in~\Sref{sec:electron}. Laser-enabled state manipulation, specifically laser cooling, play an important role in trapped atomic ion QIP experiments. Without these tools, electrical-circuit based alternatives  need to be considered along with their implications on required trap depth, low-energy electron source, electrode material, the superconducting resonator involved and a path to achieving cooling on all trap axes. Although technically challenging, these issues do not seem to preclude a hybrid system based on a trapped electron. Such a platform might offer appealing qualities such as fast entangling gates, long coherence times as well as flexible coupling topology enabled by interfacing engineered electrical circuits. 

\section{Electrical equivalent of mechanical motion}\label{sec:butterworth}
There are various systems that could, in principle, couple to a trapped charged particle. Those systems differ from the charged particle and from one another in frequency, mass, length scale, and coupling mechanism as highlighted in~\Fref{fig:mechplatforms}. With the exception of the electrical LC resonator, all other systems considered here are mechanical resonators actuated by an electromagnetic field. In order to place all of them on an equal footing we associate an electrical equivalent for each of these mechanical systems. This reduces the analysis of any of the hybrid systems into an all-electrical circuit problem. Our discussion extends the treatment in~\cite{Wineland:1975kk} where the electrical equivalent circuit of a trapped ion was derived. This could also be derived using the general framework developed by Butterworth and Van Dyke~\cite{Butterworth:1913eg,Butterworth:1914cj,VanDyke:1928gp} that associates a circuit equivalent for electrically actuated mechanical systems. We refer to the resulting electrical network as the BVD equivalent circuit.
 
\begin{figure}[!h]
	\begin{center}
		\includegraphics{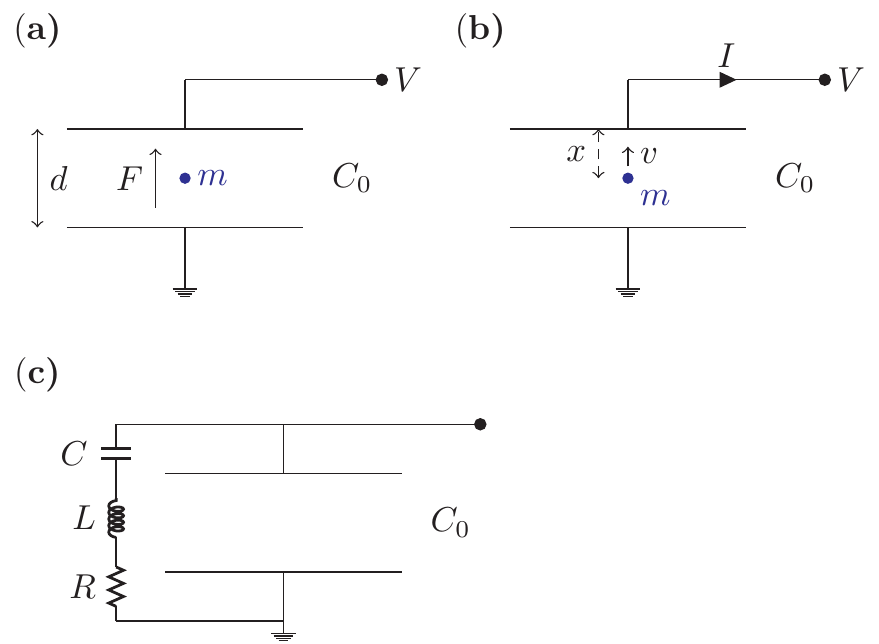}
		\caption{Simplified geometry for an electrically actuated mechanical system. \capfont{a} A mechanical system of mass $m$ is placed inside a capacitor $C_0$ that is biased at a voltage $V$. The force acting on $m$ is assumed to be proportional to the capacitor bias voltage $F=\beta V$. \capfont{b} If the mechanical system velocity is $v\neq 0$ an image current $I=\beta v$ is induced. \capfont{c} BVD equivalent circuit. The mechanical system electrical response is identical to that of a series RLC circuit connected in parallel with the capacitor $C_0$. \label{fig:bvdDiscussion}}
	\end{center}
\end{figure}

Suppose a mechanical system of mass $m$ is placed near an electrode that is biased with a voltage $V$, resulting in a force $F=\beta V$  acting on it. For simplicity, we assume the geometry in \Fref{fig:bvdDiscussion}\capfont{a}, where two electrodes form the two plates of a parallel plate capacitor, separated by a distance $d$. An important example (analyzed in~\cite{1966AmJPh..34..943S,Wineland:1975kk}) is that of a single charged particle with charge $q$ resulting in $F=qV/d$, i.e. $\beta=q/d$. In general, electrical actuation could also result from dipolar interaction, electrostriction, piezoelectricity, etc. Since microscopically these mechanisms originate from having non-zero local charge densities within the mechanical system, we lump the overall effect of the voltage with a single effective parameter, $\beta$. 

When the mass $m$ moves at a velocity $v$ (see~\Fref{fig:bvdDiscussion}\capfont{b}), it will induce a current $I=\beta v$ at the electrode. This is an immediate generalization of the single charged particle case: if it is at a distance $x$ from an electrodes it induces an image charge of $q_{\myr{image}}=qx/d$. Therefore, within the electrostatic approximation, a velocity $v=\dot{x}$ would translate into a current $I=qv/d$. The induced charges will back-act on the mass $m$ with an additional force $\Delta F$. This force, however, will be independent of $V$ and will not contribute to the induced current $I$. The effect of $\Delta F$ can therefore be lumped into a (usually but not necessarily) small change of the system's mechanical properties, e.g. its spring constant in the case of a harmonic oscillator (for a rigorous derivation see~\cite{1966AmJPh..34..943S,Wineland:1975kk}). 

Now assume that the mechanical system is harmonic, i.e. has a resonant frequency $\omega_0$ and a friction coefficient $\gamma$. If now the voltage is time varying $V(t)$, the equation of motion for the harmonic oscillator position $x$ is

\begin{equation}\label{eq:bvdeom}
	m\ddot{x}+\gamma\dot{x}+m\omega_0^2 x =\beta V(t).
\end{equation}
Using the relation $I(t)=\beta\dot{x}$ this can be rewritten as

\begin{equation}
	\frac{m}{\beta^2}\frac{dI}{dt}+\frac{\gamma}{\beta^2} I + \frac{m\omega_0^2}{\beta^2}\int^t dt'I(t')=V(t).
\end{equation}

Therefore, from the perspective of the electrical circuit, the mechanical system is equivalent to	 a series combination of resistance, inductance and capacitance, namely,
\begin{equation}
	L\frac{dI}{dt}+R I + \frac{1}{C}\int^t dt'I(t')=V(t),\\
\end{equation}
where
\begin{equation}
	L\leftrightarrow\frac{m}{\beta^2},\qquad R\leftrightarrow\frac{\gamma}{\beta^2},\qquad C\leftrightarrow\frac{\beta^2}{m\omega_0^2}, \label{eq:equivCircuitParams}
\end{equation}
and their series combination is added in parallel to the capacitance of the drive electrode $C_0$ [see~\Fref{fig:bvdDiscussion}\capfont{c}]. 

Throughout this paper, we will refer to the mechanical system and its electrical equivalent interchangeably, in order to simplify the coupling analysis. 

\section{Coupling in the strong quantum regime}\label{sec:couplinginthestrongcouplingregime}
\begin{figure}[!h]
\begin{center}
		\includegraphics{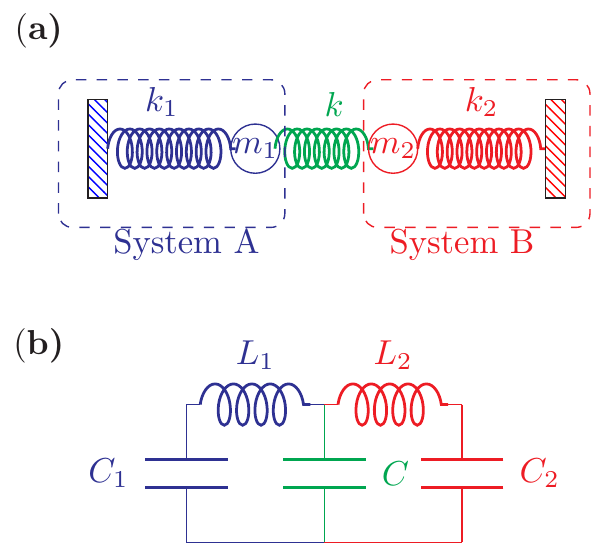}
		\caption{\capfont{a} Coupled mechanical harmonic oscillators. \capfont{b} Coupled electrical harmonic oscillators.}\label{fig:cplharmonics}
\end{center}
\end{figure}

Our general problem is concerned with two resonantly coupled harmonic oscillators (mechanical or electrical). We assume that the coupling rate $g\ll \omega_0$ so that the coupling Hamiltonian can be treated perturbatively with respect to the two harmonic oscillators hamiltonians. The Hamiltonian coupling term for two mechanical harmonic oscillators of masses $m_1,m_2$ and equal frequency $\omega_0$ by a spring of constant $k$ (\Fref{fig:cplharmonics}a) is
\begin{equation}
H_c=k x_1 x_2,
\end{equation}
where $x_{1,2}$ are the displacements of the oscillators from equilibrium. This can be rewritten in terms of a coupling rate $g$, if we express $x_i,\ i=1,2$ in terms of their respective harmonic oscillator ladder operators $x_i=\sqrt{\hbar/(2m_i\omega_0)}(\hat{a_i}+\hat{a_i}^\dagger)$ so that
\begin{equation}
	H_c=\hbar g(\hat{a_1}+\hat{a_1}^\dagger)(\hat{a_2}+\hat{a_2}^\dagger),\label{eq:coupledpendulums2} \\
\end{equation}
where
\begin{equation}
	g=\frac{k}{2\omega_0\sqrt{m_1 m_2}},\label{eq:coupledpendulums}
\end{equation} 
and $\hbar$ is the Planck constant divided by $2\pi$. 

It will be useful later to express $g$ in terms of an analog electrical system (Fig.~\ref{fig:cplharmonics}b) of two LC resonators coupled by a shunt capacitor $C$. In this case, the coupling Hamiltonian is
\begin{equation}\label{eq:couplingtwolc}
H_c=\frac{1}{C}q_1q_2,
\end{equation}
where $q_{1,2}$ are the charges on the capacitors $C_{1,2}$ respectively. The resonant frequency for each of the LC resonators is $\omega_i=1/\sqrt{L_iC_i'}$ where $C_i'=C_iC/(C_i+C)$ is the series capacitance of $C_i$ and $C$. Assuming $\omega_1=\omega_2=\omega_0$, we can rewrite~\Eref{eq:couplingtwolc} in terms of the ladder operators, $q_i=\sqrt{\hbar\omega_0C_i'/2}(a+a^\dagger)$, so that $H_c$ takes the form of~\Eref{eq:coupledpendulums2} with

\begin{equation}
	g=\frac{\omega_0}{2}\sqrt{\frac{C_1C_2}{(C_1+C)(C_2+C)}}\label{eq:capacitivecoupling}.
\end{equation}

We will be particularly interested in the \emph{strong-coupling quantum regime}, i.e. when a large number of complete energy swaps occur between the two oscillators before they significantly loose coherence: $N_{\myr{swap}}\approx \tau_{\myr{coh}}/ \tau_{\myr{swap}}\gg 1$. Here $\tau_{\myr{swap}}=\pi/g$ is the time required for a complete energy swap between the two oscillators. For a system of two harmonic oscillators $\tau_{\rm coh}$ is the average exchange time of a single energy quantum with any of the thermal baths of the oscillators. We assume that coherence is limited by energy relaxation. In reality, there are additional decoherence mechanisms which would decrease $N_{\myr{swap}}$ further and the values calculated here should be considered as an upper bound. An important case is motion dephasing of a trapped charged particle~\cite{Wineland:335673,Leibfried:2003gz}. Although the motional heating rate for trapped ions could be as low as a few quanta per second (see~\ref{apx:electronmotionheating}), trap frequency drifts, for example, could cause motional dephasing at a higher rate. Another well known source of motional decoherence is the non-linear coupling between trap axes due to trap imperfections~\cite{Wineland:335673}. Although these mechanisms could be reduced by technical means, it would be highly favorable from a practical standpoint that the coupling strength $g\ge 2\pi\times 1~\un{kHz}$, posing an additional constraint in what follows.

When expressing the above condition in terms of the lower of the two quality factors $Q$ associated with the two oscillators and the temperature $T$ of their environment we observe two regimes. At ``high'' temperatures ($k_BT\ge \hbar\omega_0$), the thermal equilibration time constant $\tau_{\myr{thermal}}=Q/\omega_0$ of the oscillators can be thought of as the $1/e$ time required to heat the mechanical oscillator from $0\ \myr{ K}$ to the surrounding temperature $T$, i.e. the time it takes to acquire an average of $(1-1/e) n_{\myr{thermal}}$ phonons where $n_{\myr{thermal}}=(e^{\frac{\hbar\omega_0}{k_BT}}-1)^{-1}\approx \frac{k_BT}{\hbar\omega_0}$ energy quanta and $k_B$ is the Boltzmann constant. Any quantum coherent phenomena will therefore be restricted to times shorter than $\tau_{\myr{coh}}=\tau_{\myr{thermal}}/n_{\myr{thermal}}$, roughly the time required to absorb one phonon at the rate of thermal equilibration. At ``low'' temperatures ($k_B T\leq \hbar \omega_0$) the equilibrated oscillator contains one phonon or less on average and therefore $\tau_{\myr{coh}}=Q/\omega_0$. The strong quantum regime condition therefore translates to
	
	\begin{equation}\label{eq:gQ}
		N_{\myr{swap}}\approx\frac{gQ}{\pi(n_{\myr{thermal}}+1)\omega_0}\gg 1.
	\end{equation}
At typical liquid helium temperatures of $\sim 4~\un{K}$, $k_BT/\hbar=2\pi\times 83~\un{GHz}$ so for frequencies below $83~\un{GHz}$ we require
	\begin{equation}\label{eq:gQat4K}
		N_{\myr{swap}}\approx\frac{gQ}{2\pi\times 262~\myr{GHz}}\gg 1.
	\end{equation}
For dilution-refrigerator temperatures of $\sim 50~\un{mK}$ for example, $k_BT/\hbar=2\pi\times 1~\un{GHz}$ so for frequencies below $1~\un{GHz}$ we require
	\begin{equation}\label{eq:gQat50mK}
		N_{\myr{swap}}\approx\frac{gQ}{2\pi\times 3.3~\myr{GHz}}\gg 1.
	\end{equation}
	
The inequalities in~\eref{eq:gQ}-\eref{eq:gQat50mK} introduce stringent constraints both on the coupling strength $g$ and the Q-factors involved. The need for high Q-factors accounts for the reason why superconducting circuits, which often have high Q-factors, naturally arise in the context of hybrid systems, as will be seen in the next section.  
	
If the two oscillators have different eigen-frequencies ($\omega_1\neq\omega_2$) their weak off-resonant coupling could be brought into a strong effective resonant coupling by modulating one or more of the system parameters by a fraction $0<\eta<1$, at the difference frequency, $\omega_1-\omega_2$, usually at the expense a lower coupling rate.  For example, if the two mechanical oscillators in~\Fref{fig:cplharmonics}\capfont{a} have different resonant frequencies, they can still be coupled by modulating the spring constant $k$ at the difference frequency. The expression for the coupling rate  in~\Eref{eq:coupledpendulums2} generalizes to $g=\eta k/(4\sqrt{\omega_1\omega_2m_1m_2})$. Therefore, the coupling strength is reduced by $\eta/2$, where $\eta$ is typically at the $0.05-0.2$ range to avoid non-linear behavior of the coupling spring. Since large coupling rates are critical we concentrated on resonant oscillators in the above discussion and in what follows. For details of parametric coupling schemes in the context of hybrid systems involving ions see~\cite{Kielpinski:2012hw,Daniilidis:2013kw,DeMotte:2015wy,Kafri:2015uf}. 
	
\section{Trapped charged particle coupled to an electrical resonator}\label{sec:supercondresonator}
\begin{figure}[!h]
	\begin{center}
		\includegraphics{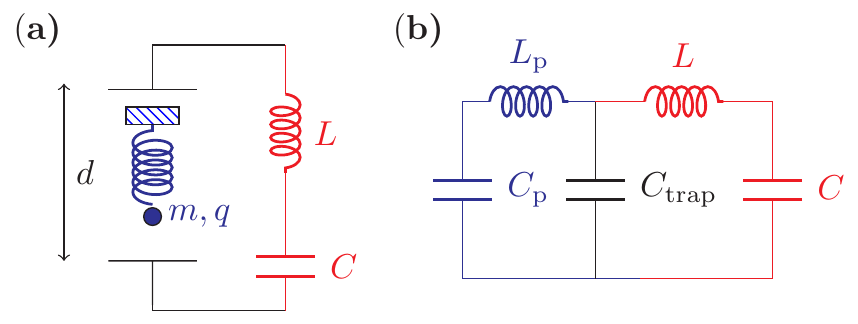}
		\caption{\capfont{a} A simplified picture of a trapped particle coupled to an LC resonator. \capfont{b} The corresponding electrical BVD equivalent circuit. The trap capacitance $C_{\myr{trap}}$ in \capfont{b} is formed by the two equivalent parallel plates which are a distance $d$ from one another in \capfont{a}.\label{fig:Heinzen}}
	\end{center}
\end{figure}

The first hybrid system we consider is that of a trapped charged particle coupled to an electrical resonator, following \cite{Heinzen:1990zz} (see also \cite{Daniilidis:2013jz}). Schematically, a point particle of mass $m$ and charge $q$ is elastically bound by a trap, here modeled by a spring (see~\Fref{fig:Heinzen}). If the particle is placed between the two plates of a capacitor, any voltage difference $V$ between the plates would result in a force $F=\alpha q V/d$ acting on it, where $d$ is the distance between the plates and $\alpha$ is a unit-less geometric factor ($\alpha=1$ for a parallel plate capacitor with infinite plate areas). Using the equivalent electrical circuit (\Eref{eq:equivCircuitParams}), the mass $m$, charge $q$ and resonant radial frequency $\omega_0$, translate into an effective inductance $L_{\myr{p}}$ and capacitance $C_{\myr{p}}$:

\begin{equation}
L_{\myr{p}}=\frac{md^2}{\alpha^2 q^2},\qquad C_{\myr{p}}=\frac{1}{L_{\myr{p}}\omega_0^2}.\label{eq:cionlion}
\end{equation}
Therefore, the hybrid system composed of a harmonically confined charged particle and resonator is equivalent to a lumped element LC circuit ($L_{\myr{p}},C_{\myr{p}}$) shunted by the trap capacitance $C_{\myr{trap}}$, and coupled to the electrical resonator, as shown in \Fref{fig:Heinzen}\capfont{b}. From~\Eref{eq:capacitivecoupling} and assuming $C\gg C_{\rm trap}$ for maximal coupling,  we get:

\begin{equation}
g=\frac{\omega_0}{2}\sqrt{\frac{C_{\myr{p}}}{C_{\myr{trap}}}}=\frac{\alpha q}{2d}\frac{1}{\sqrt{mC_{\myr{trap}}}}.\label{eq:ioncoil}
\end{equation}
This coupling can be increased by trapping more than one charged particle. If $N_p$ particles are trapped and form a Wigner crystal, their common mode motion can be treated as that of a single particle with a charge of $N_p q$ and a mass of $N_p m$. From \Eref{eq:ioncoil} it follows that $g\propto \sqrt{N_{\rm p}}$.   For very small traps however, $N_{\rm p}$ will be limited by the Coulomb repulsion between the charges.
	
\begin{table*} 
		\begin{ruledtabular}
			\begin{tabular}{llllll}
				Particle      & Mass, $m$      & Trap frequency, $\omega_0$ & Coupling strength, $g$ & $Q_{\myr{min}}(4K)$ & $Q_{\myr{min}}(50~\un{mK})$ \\
				\hline 
				electron      & $m_e$          & 1.3 GHz                    & 1.2 MHz                & $4\times 10^5$      & $7\times 10^3$              \\
				$^9$Be$^+$    & $9\times m_p$  & 10 MHz                     & 9 kHz                  & $56\times 10^6$     & $7\times 10^5$              \\
				$^{24}$Mg$^+$ & $24\times m_p$ & 6 MHz                      & 6 kHz                  & $92\times 10^6$     & $1.1\times 10^6$            \\
				$^{40}$Ca$^+$ & $40\times m_p$ & 4.7 MHz                    & 4 kHz                  & $119\times 10^6$    & $1.5\times 10^6$            \\
				$^{88}$Sr$^+$ & $88\times m_p$ & 3.2 MHz                    & 3 kHz                  & $176\times 10^6$    & $2\times 10^6$
			\end{tabular}
		\end{ruledtabular}
			\caption{Coupling strengths of different trapped charged particles coupled to an electrical resonator. The mass of the proton and the electron are $m_p$ and $m_e$ respectively. We assume the geometry in \Fref{fig:Heinzen}, with $d=50\ \mu$m, $C_{\myr{trap}}=50\ \myr{fF}$ and $\alpha=1$, and use~\Eref{eq:ioncoil} to calculate $g$. The table states a lower bound for the required Q-factors, namely $Q$ corresponding to $N_{\myr{swap}}=1$. Actual Q-factors should be at least an order of a magnitude greater to comfortably satisfy inequality~\Eref{eq:gQ}. These estimates are consistent with~\cite{Daniilidis:2009ga} where $600~\un{Hz}$ coupling strength was estimated for $^{40}$Ca$^+$ in a $1~\un{MHz}$ trap with $2.5~\un{pF}$ trap capacitance, $d=50~\mu\un{m}$ and $\alpha=1$.\label{tbl:ioncoil}}
\end{table*}
	
Based on~\Eref{eq:gQ}, \Tref{tbl:ioncoil} summarizes the constraints on the Q-factor of the electrical resonator required to be in the strong-coupling quantum regime for various charged particles. These should be compared to experimentally attainable values for lumped-element superconducting resonators that are typically in the range of $Q\sim 10^4-10^5$ and in some cases up to $10^6$, mostly limited by dielectric losses~\cite{Wenner:2011co,Geerlings:2012ju}. Since the required $Q$ is greater than these values, achieving strong coupling of an ion to a superconducting resonator at $4~\un{K}$ does not seem feasible. In fact, the only two candidates from \Tref{tbl:ioncoil} that stand out in terms of reasonable Q-factors are \Be ($Q\gg 7\times 10^5$ at $50~\un{mK}$) and electrons ($Q\gg 4\times 10^5$ at $4~\un{K}$ and $Q\gg 7\times 10^3$ at $50~\un{mK}$). For \Be it would require incorporating atomic ion trapping technology into a dilution refrigerator, the discussion of which is beyond the scope of this paper and can be found elsewhere~\cite{DeMotte:2015wy}. We discuss the prospects of electron coupling in the last part of the paper. Our estimates are compatible with previous results~\cite{Daniilidis:2009ga,Daniilidis:2013kw}.	
	
\begin{figure}[!h]
	\begin{center}
		\includegraphics{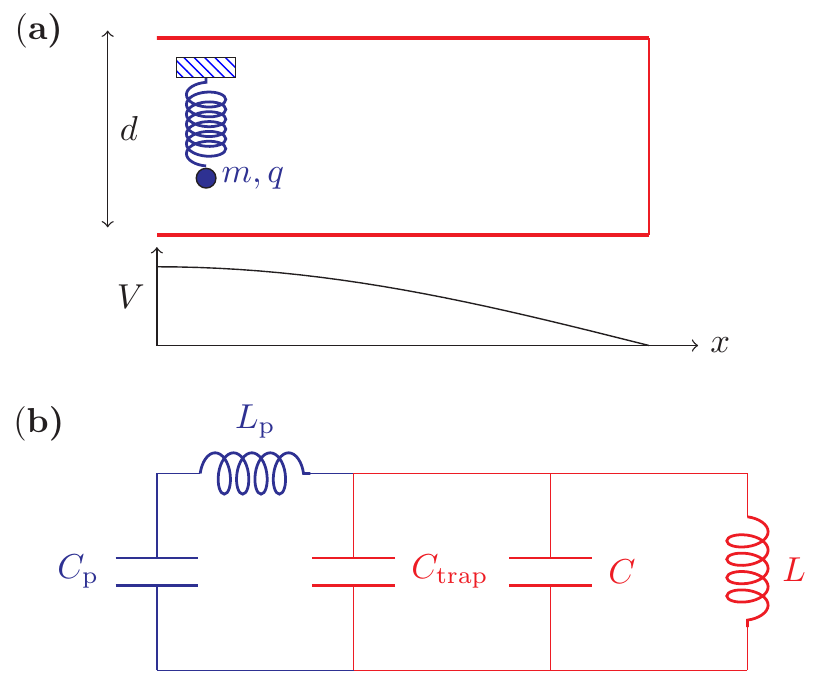}
		\caption{\capfont{a} A simplified picture of a trapped ion coupled to a transmission line resonator. The ion is trapped close to the voltage anti-node of a short circuited quarter wave resonator. \capfont{b} The corresponding electrical equivalent circuit. The ion is replaced with its equivalent series capacitance $C_p$ and inductance $L_p$ while the resonator is replaced with its equivalent lumped element representation formed by a parallel $LC$ resonator. Additional capacitance due to trap electrodes is represented by $C_{\myr{trap}}$.\label{fig:ionandquarterwave}}
	\end{center}
\end{figure}	
	
In the above discussion we considered only lumped-element electrical resonators. A different approach would be to use low frequency transmission line resonators.  Those can be simpler to fabricate and could potentially have higher quality factors. As an example,~\Fref{fig:ionandquarterwave}\capfont{a} shows a simple geometry where an ion is trapped close to the voltage anti-node of a quarter-wave resonator. Near resonance, the transmission line resonator is equivalent to a parallel $LC$ circuit (see \Fref{fig:ionandquarterwave}\capfont{b}) with effective capacitance $C=\pi/(4\omega_0Z_0)$ and inductance $L=1/(\omega_0^2 C)$ where $\omega_0$ is the resonance frequency and $Z_0$ the characteristic impedance of the transmission line~\cite{Pozar:2011to}. The coupling strength is calculate, as before, using the electrical equivalent circuit:

\begin{equation}
g=\frac{\omega_0}{2}\sqrt{\frac{C_{\myr{p}}}{C+C_{\myr{trap}}}}.\label{eq:ionquarterwave}
\end{equation}

The main concern is that the effective capacitance $C$ of these resonator modes is very large. For a typical $Z_0=50~\un{ohms}$ transmission line and $\omega_0=2\pi\times 10~\un{MHz}$, $C\sim 250~\un{pF}$. The coupling strength $g$ will therefore degrade by a factor of $\sim 70$ as compared to the numbers in \Tref{tbl:ioncoil}, requiring, for example, a quality factor satisfying $Q\gg 4\times 10^9$ for \Be at 4 K. This number exceeds the best quality factors for such resonators, having $Q\sim 10^7$ at $10~\un{MHz}$~\cite{Erickson:2014ep}. Moreover, our estimate for $g$ is an upper bound since in a real geometry, the field lines at the voltage anti-node of the resonator will differ from those of an ideal parallel plate capacitor. For those reasons, our analysis has focused on coupling the charged particle to a lumped-element electrical oscillator, where the same resonant frequency can usually be achieved with significantly less overall capacitance.

\section{Coupling to macroscopic mechanical resonators}
To circumvent the limitations on attainable Q-factors of superconducting devices, it has been suggested to try and couple an ion directly to a high-Q macroscopic mechanical object using electro-static coupling~\cite{Heinzen:1990zz,Wineland:335673,Hensinger:2005kc,Hunger:2011eo,Daniilidis:2013jz} or piezoelectricity \cite{Heinzen:1990zz,Taylor:jmsWUKDC}. 

\subsection{Electrostatic coupling to a nano-mechanical membrane}
\begin{figure}[!h]
\begin{center}
	\includegraphics{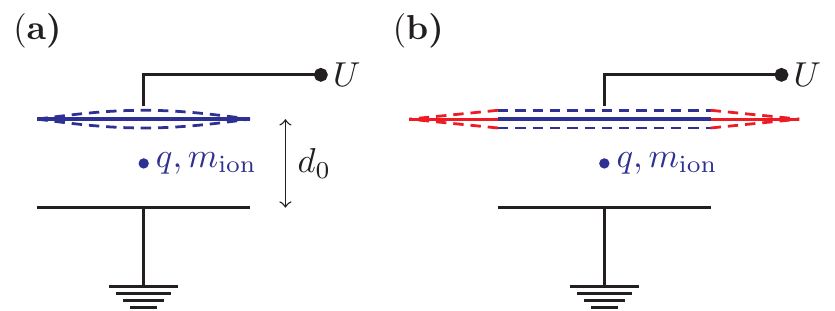}
	\caption{Electrostatic coupling of a trapped ion (charge $q$ and mass $m_{\myr{ion}}$) to a nearby rectangular nano-mechanical membrane biased by a voltage $U$. The ion is assumed to be trapped at a height $d_0/2$ above a surface trap, that is dc grounded with respect to the membrane, suspended above the ion (for simplicity the trap rf electrodes are omitted). \capfont{a} A membrane (blue) is clamped at its rim, allowing for a sinusoidal fundamental mode as in~\cite{Yu:2014ky}. \capfont{b} A membrane (blue) is attached by thin wires (red), allowing for a center of mass fundamental mode as in~\cite{Reinhardt:2016ef,Norte:2016eq}.}\label{fig:ionmembrane}
\end{center}
\end{figure}

Commercial nano-mechanical membrane resonators can have high quality factors of over $10^7$ at $300~\un{mK}$~\cite{Zwickl:2008fa}. Recent advances in membrane fabrication~\cite{Yu:2014ky,Teufel:2016eq,Reinhardt:2016ef,Norte:2016eq,Tsaturyan:2016tt} resulted in quality factors as high as $10^8$, even at room temperature. If such a membrane is metalized on one side, and biased with a voltage $U$, it would electro-statically couple to an ion trapped near its surface. To get an estimation for the coupling, we assume the simple geometries in~\Fref{fig:ionmembrane}. In both cases, the coupling Hamiltonian is
\begin{equation}
H=\alpha qU\frac{z_iz_m}{d_0^2},
\end{equation}
where $z_i,z_m$ are the displacements of the ion $z$-motion and the membrane, respectively, $d_0$ is the distance between the membrane and the bottom electrode of the ion trap and $\alpha$ is a geometric factor as in~\Sref{sec:supercondresonator}. For the geometries considered here $0.5\le \alpha\le 1$ and we assume $\alpha=1$ to get an upper bound for $g$. As in~\Eref{eq:coupledpendulums2}, we can derive the coupling strength
\begin{equation}
g=\frac{\alpha qU}{2d_0^2\omega_0\sqrt{m_{\myr{ion}}M}},
\end{equation}
where $M$ is the membrane mode mass and $\omega_0$ its resonant frequency. We assume that $d_0=100\ \mu\un{m}$ and the ion is trapped midway between the membrane and the trap. For a SiN membrane~\cite{Yu:2014ky} with dimensions $500~\mu\un{m}\times 500~\mu\un{m}$ coupled to a \Be ion, we get $g/2\pi\sim 0.24~\un{Hz}$ at $U=1~\un{V}$  bias and a resonant frequency $\omega\sim 2\pi\times 1~\un{MHz}$. Combined with an assumed quality factor of $2\times 10^8$ such a device does not satisfies the strong quantum criteria at $T=50~\un{mK}$ since $gQ/2\pi\sim 0.048~\un{GHz}$ (see~\Eref{eq:gQat50mK}). For a suspended trampoline membrane~\cite{Reinhardt:2016ef,Norte:2016eq} with dimensions $100~\mu\un{m}\times 100~\mu\un{m}$ coupled to a \Be ion, we get $g/2\pi \sim 12~\un{Hz}$ and $gQ/2\pi\sim 1.2\  \un{GHz}$ at $U=1~\un{V}$ and a resonant frequency of  $\omega\sim 2\pi\times 140~\un{kHz}$. The latter nearly enters the strong quantum regime for $T=50~\un{mK}$. However, taking into account ion heating rates still make this scheme unfavorable, since ion motional heating rate and motional dephasing would typically exceed $g$. 

The coupling can be made stronger by increasing the bias voltage $U$ at the expense of changing the trapping potential, the ion position and possibly the trapping stability. Even with the $U=1~\un{V}$ assumed above, the equilibrium position of, say a \Be ion in a $10~\un{MHz}$ harmonic trap, would move by $\sim 7~\mu\un{m}$. This might be mitigated by adding additional electrodes which compensate for the static voltage bias effect of the membrane (e.g. see~\cite{Daniilidis:2013jz}). Those electrodes, however, might shield some of the trapping field and need to be taken into account when estimating the ion trapping potential. In addition, a more careful estimation of $g$ would take the membrane mode shape and finite size into account. Finally, adding an electrode to a membrane might decrease its $Q$-factor. Previous experiments~\cite{Andrews:2014kj} with lower quality factors ($Q\sim 10^6$) showed that metalization of the membrane was not the limiting factor. Whether or not this is also true for the case of $Q\sim 10^8$ would need to be tested experimentally. 

\subsection{Piezoelectric coupling to an acoustic resonator}
A piezoelectric resonator is an acoustic resonator made from piezo-electric materials and can therefore be excited using external electric fields~\cite{Cady:279552}. Quartz resonators have been optimized for stable frequency operation and are therefore natural candidates for ion coupling, despite being relatively massive. A different plausible candidate is GaN-nanobeams that have low masses.
	
\begin{figure}[!h]
		\begin{center}
		\includegraphics{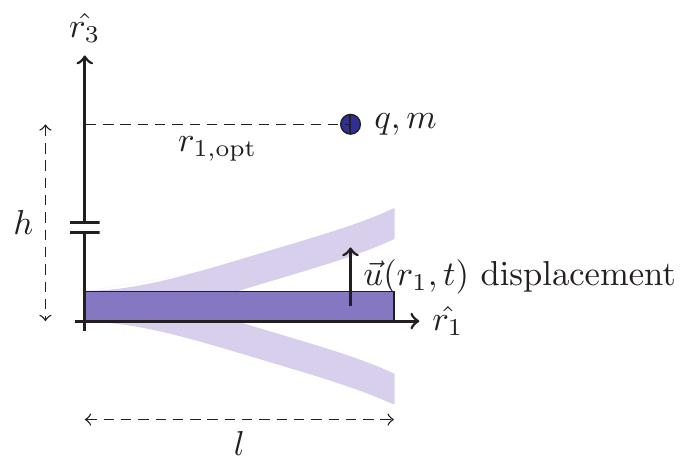}
		\caption{Piezo coupling between an ion of mass $m$ and charge $q$ to a nano-beam. The ion is held at a height $h$ above a beam of length $l$ by a Paul trap (not shown). The geometry shown is not to scale since $h\gg l$ (see~\Sref{sec:nanobeam}). Harmonic motion about the trap center generates an alternating electric field which drives the mechanical flexure mode of the beam (light blue) via the piezoelectric effect. The ion position $r_{1,\myr{opt}}$ maximizes the coupling and is close to but smaller than the beam length $l$, due to edge effects.}
		\label{fig:nano}
		\end{center}
	\end{figure}

To estimate the coupling strength, we start by considering the geometry shown in~\Fref{fig:nano}. An ion is trapped at a distance $h$ above a GaN nano-beam. Such an arrangement can be achieved, for example, by bringing a surface ion trap~\cite{Chiaverini:2005ur,Seidelin:2006hg} or a stylus ion trap~\cite{Maiwald:2009fp,Arrington:2013il} close to the beam. The main challenge would seem to be to compensate for electric fields from stray charges on the dielectric beam due to its close proximity. We assume throughout that those are compensated for. When such a beam undergoes small oscillations, the position of each point in the beam can be written as $\vec{r}+\vec{u}(\vec{r},t)$ where $\vec{r}=(r_1,r_2,r_3)$ is the equilibrium position and $\vec{u}=(u_1,u_2,u_3)$ is the time-dependent displacement from equilibrium. In a flexure acoustic mode $\vec{u}$ is along the $\hat{r}_3$ direction and its spatial dependence is restricted to the first component of $\vec{r}$ (see~\Fref{fig:nano}). Moreover, the dependence on time and spatial coordinates can be separated, i.e. $\vec{u}(r_1,t)=a(t)\vec{s}(r_1)$, where $\vec{s}(r_1)=(0,0,s_3(r_1))$ is the mode shape (unit-less) and $a(t)$ is its amplitude. The acoustic oscillation can therefore be reduced to a one-dimensional harmonic oscillator $a(t)$ with frequency $\omega_0$, effective mode-mass $M$ and effective spring constant $K$:
	\begin{subequations}
	\begin{eqnarray}
		M\ddot{a}&=&-K a,\\
		M&=&\rho \int_{V}d^3r \abs{s}^2,\\
		K&=&E \int_{V}d^3r \abs{\frac{ds}{dr_1}}^2,
	\end{eqnarray} 
	\end{subequations}
	where $\rho$ is the material density, $V$ is the volume of the beam and $E$ is its Young's modulus. 
	
	The harmonic motion of the ion can couple to the beam acoustic mode via piezoelectricity. A simplified model of the beam piezoelectric material is that of an ionic lattice. When the beam is at rest, the electric fields generated by the positive and negative charges inside it ideally cancel each other. If, however the ions are displaced from equilibrium non-uniformly\footnote{A uniform displacement of all the ions cannot generate bulk polarization.} the beam will exhibit a bulk polarization $P$ that can interact with the electric field of the ion. Such a polarization therefore, depends linearly on the strain tensor composed of all the partial derivatives of the displacement components $\partial_i u_j\equiv \partial u_j/\partial r_i$ for $i,j\in \{1,2,3\}$. Since the strain tensor is symmetric, this linear relation can be written as $\vec{P}=eu'$ where $e$ is the $3\times 6$ matrix of piezo coefficients (in units of Cm$^{-2}$) and $u'$ represents strain in Voigt notation $u'=(\partial_1 u_1,\partial_2u_{2},\partial_3u_{3},\partial_2u_{3}+\partial_3u_2,\partial_3u_{1}+\partial_1u_3,\partial_1u_{2}+\partial_2u_1)$. This bulk polarization will in turn be influenced by the ion electric field $\vec{E}_{\myr{ion}}$. The coupling constant between the ion motion along the $i$-th axis and the piezo-electric beam is 
	
	\begin{equation}\label{eq:piezo_coupling_overlap_integral}
		g_i=\frac{\int_V d^3r \partial_i \vec{E}_{ion}es'}{2\omega_0\sqrt{M m_{ion}}},\quad i=1,2,3.
	\end{equation}
Here we used the assumption that $\vec{u}=a(t)\vec{s}(r_1)$ and $s'$ is defined in the same manner as $u'$.

The expression in~\Eref{eq:piezo_coupling_overlap_integral} is general and not particular to a beam geometry. While the denominator is the standard term we encountered for two coupled mechanical oscillators (see~\Eref{eq:coupledpendulums2}), the numerator is a rather involved overlap integral. In order to appreciate its complexity, we write its integrand in explicit matrix form: 

\begin{equation}
\partial_i \left(\begin{smallmatrix}
E_{\myr{ion},1},E_{\myr{ion},2},E_{\myr{ion},3}
\end{smallmatrix}\right)\left(\begin{smallmatrix}
e_{1,1}& \ldots & e_{1,6}\\
\ldots& 				&		\\
e_{3,1}&\ldots			& e_{3,6}\\
\end{smallmatrix} \right)
\left(\begin{smallmatrix}
\partial_1s_{1}\\
\partial_2s_{2}\\
\partial_3s_{3}\\
\partial_2s_{3}+\partial_3s_{2}\\
\partial_3s_{1}+\partial_1s_{3}\\
\partial_1s_{2}+\partial_2s_{1}\\
\end{smallmatrix}\right)
.\label{eq:piezooverlap}
\end{equation}
This integrand can be understood as a dipole-dipole energy density. To see this, notice that since the field of the ion is that of a monopole, its spatial derivative $\partial_i \vec{E}_{\myr{ion}}$ is equivalent to a dipole field aligned along the $i$-th axis, $\hat{i}$. We may therefore rewrite~\Eref{eq:piezo_coupling_overlap_integral}-\eref{eq:piezooverlap} in terms of an integral over an effective dipole-dipole interaction:

\newcommand{\pion}{\vec{p}_{\myr{ion}}}
\newcommand{\vP}{\vec{P}}
\begin{equation}\label{eq:piezo_as_dipole_dipole}
g_i=\frac{1}{4\pi\hbar\bar{\epsilon}} \int_V d^3r \frac{3(\pion\cdot \hat{r})(\vP\cdot \hat{r})-\pion\cdot\vP}{r^3},
\end{equation}
where
\begin{subequations}
	\begin{eqnarray}
	\pion&=q\sqrt{\frac{\hbar}{2m_{\myr{ion}}\omega_0}}\hat{i},\\
	\vP&=es'\sqrt{\frac{\hbar}{2M\omega_0}},
	\end{eqnarray}
\end{subequations}
and we use $\bar{\epsilon}=(\epsilon_0+\epsilon_{\myr{dielectric}})/2$ since the field of the ion inside the piezoelectric material can be approximated as that of an ion in vacuum, with the dielectric constant of vacuum $\epsilon_0$ replaced by $\bar{\epsilon}$, the average of the vacuum and dielectric dielectric constants~\cite{Jackson:490457}. 

A priori, the overlap integral in the numerator of \Eref{eq:piezo_coupling_overlap_integral} should not be expected to be large. The piezo-electric coefficient matrix $e$ is a material property, while the mode shape $\vec{s}$ is a result of both geometry and material constraints. Those impose a polarization density $\vP$ which need not necessarily align with $\pion$. We next perform a calculation for two specific piezo-electric resonators in order to demonstrate this difficulty. We use~\Eref{eq:piezo_coupling_overlap_integral} and~\Eref{eq:piezo_as_dipole_dipole} interchangeably. 
	
\subsection{Ion coupled to a GaN nanobeam}\label{sec:nanobeam}
\begin{figure}[!h]
\begin{center}
		\includegraphics[scale=0.5]{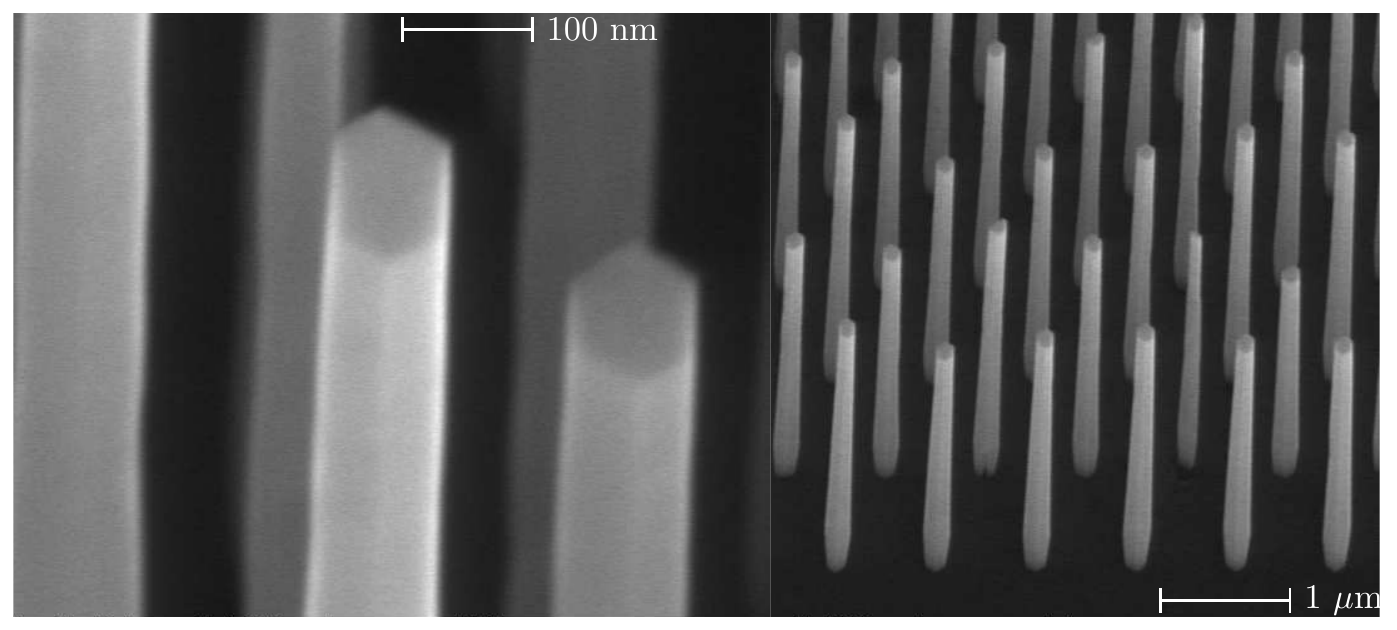}
	\caption{SEM microscopy of GaN nanobeams with hexagonal cross section. Gallium nitride nano-beams photos courtesy of K. Bertness at NIST, Boulder, CO 80305, USA.}\label{fig:fromrogers}
\end{center}
\end{figure}

Figure~\ref{fig:fromrogers} shows an image of Gallium Nitride (GaN) nano-beams. A single beam, clamped at one end, can resonate in a flexure mode \cite{Cleland:1624966} with a resonance frequency of $\omega_0=\sqrt{\beta a^2 E/\rho l^4}$. Here, $a$ is the cross-section radius, $l$ is the beam length, $E$ is its Young's modulus, $\rho$ is its density, and $\beta$ is a numerical factor ($3.09$ for a circular cross section, $2.57$ for a hexagonal cross-section\footnote{for a hexagon, the radius is defined to be that of the smallest circle enclosing it.}). 

We can estimate an upper limit on the coupling rate based on \Eref{eq:piezo_coupling_overlap_integral} and using the simplified geometry in~\Fref{fig:nano}:
	\begin{equation}
		g=\frac{q\tilde{e}A}{4\pi\bar{\epsilon} h^3 \omega_0\sqrt{Mm_{ion}}} f(h/l),\label{eq:GaNnanobeam}
	\end{equation}
where $f$ is a unitless geometric factor depending on the $h/l$ aspect ratio, $A$ is the cross section area, $\tilde{e}$ is the largest element of the $3\times 6$ GaN piezo-coefficient matrix and $\bar{\epsilon}$ is the average of its dielectric constant and that of vacuum. The ion position along the beam $r_{1,\myr{opt}}$ is chosen so as to maximize the coupling. It turns out $r_{1,\myr{opt}}\sim 0.6~ l$ due to edge effects.

\begin{figure}[!htbp]
		\begin{center}
			\includegraphics{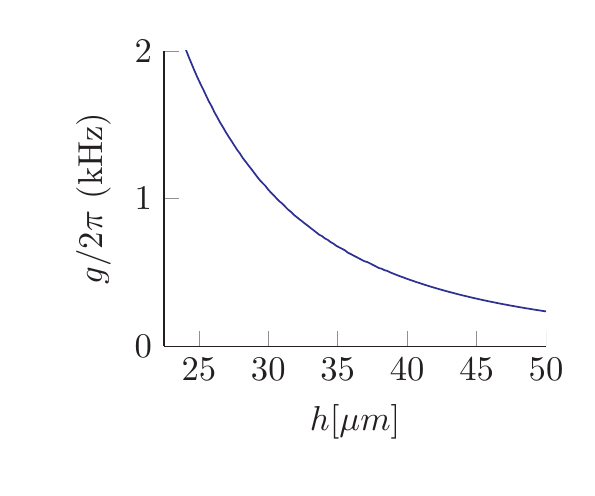}
		\caption{Ion to GaN nanobeam piezoelectric coupling strength $g$ vs. ion height $h$ above the beam. The beam cross section is as in \Fref{fig:fromrogers}. The geometry is as in Fig.~\ref{fig:nano} with $l=15\ \mu\un{m}$, $E=3\times 10^{11}~\myr{kg\ m}^{-1}s^{-2}$, $\rho=6.15\times 10^4~\myr{kg\ m}^{-3}$, $\tilde{e}=0.375~\un{Cm}^{-2}$ (the strongest piezo coefficient of GaN), $\bar{\epsilon}=5\epsilon_0$, $\epsilon_0$ being the vacuum permittivity. The beam flexure mode frequency is $868~\un{kHz}$.\label{fig:nanocpl}}
		\end{center}
\end{figure}

Figure \eref{fig:nanocpl} shows the coupling coefficient as a function of ion height $h$. At an experimentally attainable height of $h=50\ \mu\un{m}$, beam length $l=15~\mu\un{m}$ and frequency $\omega_0=2\pi \times 868~\un{kHz}$, the coupling strength is $g=2\pi\times 235~\myr{Hz}$. Even for a relatively high quality factor beam of $Q=6\times 10^4$~\cite{Rogers2007}, the product $gQ/2\pi=1.4\times 10^7~\un {Hz}$ whereas the strong quantum regime requires $gQ/2\pi\gg 2.6\times 10^{11}~\un{Hz}$ at $4\ K$ and $gQ/2\pi \gg 3.3\times 10^9~\un{Hz}$ at $50~\un{mK}$ (\Eref{eq:gQ}).
	
Based on \Eref{eq:GaNnanobeam}, the coupling to materials other than GaN can be estimated. Another notable material is Lithium Niobate where the strongest of the piezo-electric coefficients is an order of a magnitude larger than for GaN, with the other parameters reasonably close to GaN~\cite{Gualtieri1994}. That, however, would still have a $gQ$ factor which is below our criteria ($N_{\myr{swap}}\sim 10^{-4}$ at $4~\un{K}$), and even that estimate assumes a high-Q Lithium Niobate resonator, which has yet to be demonstrated. Another approach would be to use beams with higher quality factors that are close to $10^6$, for example silicon nitride~\cite{Verbridge:2006ix} doubly clamped beams or other resonators (see tables 1 and 2 in~\cite{Poot:2012fh}). However, since these resonators are not made from piezoelectric material, it would require incorporating piezoelectric material into the beam while maintaining the high quality factors.

\subsection{Ion coupled to a quartz resonator}
Recent work with quartz bulk acoustic resonators at both $4~\un{K}$ and tens of  millikelvin temperatures demonstrated quality factors of up to $7.8\times 10^9$ and might therefore be useful as part of  a hybrid quantum system~\cite{Maxim2011,Maxim2012,Galliou:2013br,Goryachev:2013gl,2014ApPhL.105o3505G}. Conveniently, the resonance frequencies of these devices are compatible with those of trapped ions, i.e. in the $5-15$ MHz range.
	
\begin{figure}[!h]\begin{center}
		\includegraphics[scale=1]{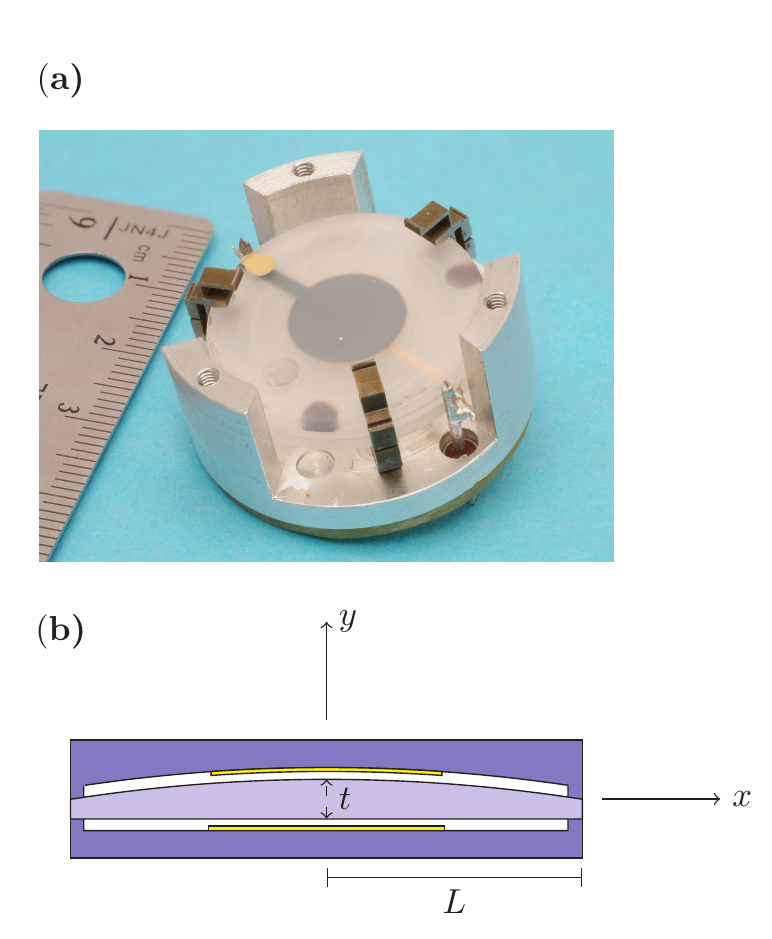}
		\caption{High-$Q$ quartz bulk acoustic resonator. \capfont{a} Photo of a resonator. Device courtesy of Serge Galliou, FEMTO-ST institute, 25000 Besan\c{c}on, France. \capfont{b} Schematic cross section. Quartz resonator of thickness $t$ is shown by the light blue fill. Quartz holders (dark blue fill) clamp the resonator at its rim. The resonator is sandwiched between two metallic electrodes forming the actuating capacitor (yellow fill). Thickness of the electrodes as well as the gap between the quartz resonator and the quartz holders are exaggerated for clarity. The modes with highest Q-factor can be described by standing waves approximately along the $y$ axis, with resonant frequencies of $f_n\approx n\frac{v_s}{2t}=n\times 3.38~\myr{MHz}$ where $v_s=6757~\myr{m/s}$ is the speed of sound and $n$ is the mode number.}\label{fig:bvapic}
	\end{center}
\end{figure}
	
\begin{figure}[!h]
		\begin{center}
		\includegraphics{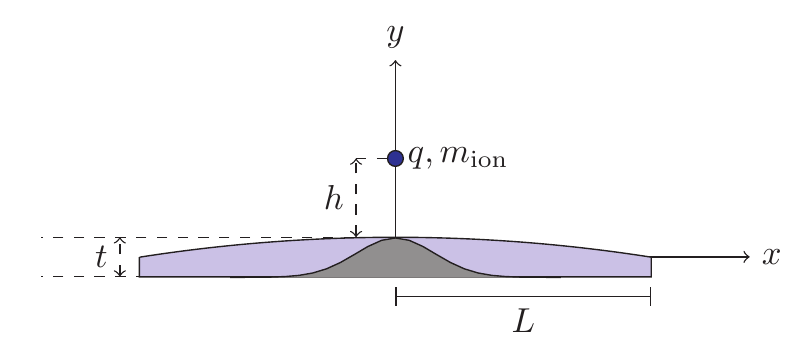}
		\caption{Basic geometry for ion-to-quartz resonator coupling. An ion of mass $m_{\myr{ion}}$ and charge $q$ is hovering at a distance $h=50\ \mu\un{m}$ (exaggerated) above a disk of radius $L=6.5~\un{mm}$ and thickness $t=1~\un{mm}$. The Gaussian radial profile of the acoustic mode is shown in gray. The ion motion generates an oscillating electric field that actuates the acoustic modes via piezoelectricity}\label{fig:ionabovebva}
		\end{center}
	\end{figure}

A BVA resonator (Bo\^itier \'a Vieillissement Am\'elior\'e, Enclosure with Improved Aging), is a quartz resonator designed for high-Q clock oscillators~\cite{Besson:1995va}. The resonator described here is formed from a disk of $L=6.5~\un{mm}$ radius and $t=1~\un{mm}$ thickness mechanically clamped at its rim (see \Fref{fig:bvapic}). The mechanical motion of the disk is actuated by placing the disk between the two plates of a capacitor. The origin of the high Q-factors becomes apparent when considering the mechanical displacement profiles of one family of its acoustic modes~\cite{Stevens:1986im}:
	\begin{equation}
		\vec{s}(x,y,z)=e^{-(x^2+z^2)/2\sigma^2}\sin(k_n y)\hat{s}.\label{eq:bvamode}
	\end{equation}
Here an acoustic standing wave is formed along the unit vector $\hat{s}=(0.226, 0.968, 0.111)$ which is approximately along the $\hat{y}$ axis (see~\Fref{fig:ionabovebva}). The mode $k$-vector satisfies $k_n t=n\pi,\ n=3,5,\ldots$ and has a radial Gaussian profile, with $\sigma\sim 1~\myr{mm}<L$. This is very similar to the standing wave formed in a Fabry-P\'erot optical cavity.  The acoustic mode is therefore well protected from dissipation through the rim, where the disk is clamped. Other acoustic mode families are not considered here since they exhibit lower quality factors~\cite{Galliou:2013br}. This is also the reason why we do not consider the fundamental $n=1$ mode of \Eref{eq:bvamode}.
	
An ion can be coupled to the quartz resonator by trapping it a distance $h=50\ \mu\un{m}$ from the surface, as shown in~\Fref{fig:ionabovebva}. Calculating the coupling strength can be accomplished using \Eref{eq:piezo_coupling_overlap_integral} and considering the acoustic mode shape (see~\Eref{eq:bvamode}). An upper bound, which does not take into account the relative angle between the derivative of the field of the ion and the polarization of the bulk, yields $g\sim 2\pi\times 1~\myr{kHz}$. This is calculated by applying the Cauchy-Schwartz inequality to the integrand in~\Eref{eq:piezooverlap} of the overlap integral in \Eref{eq:piezo_coupling_overlap_integral}. Combined with the high quality factors involved ($Q\sim 10^9$) this yields $gQ/2\pi\sim 10^{12}~\un{Hz}$. 

\begin{figure*}
			\includegraphics{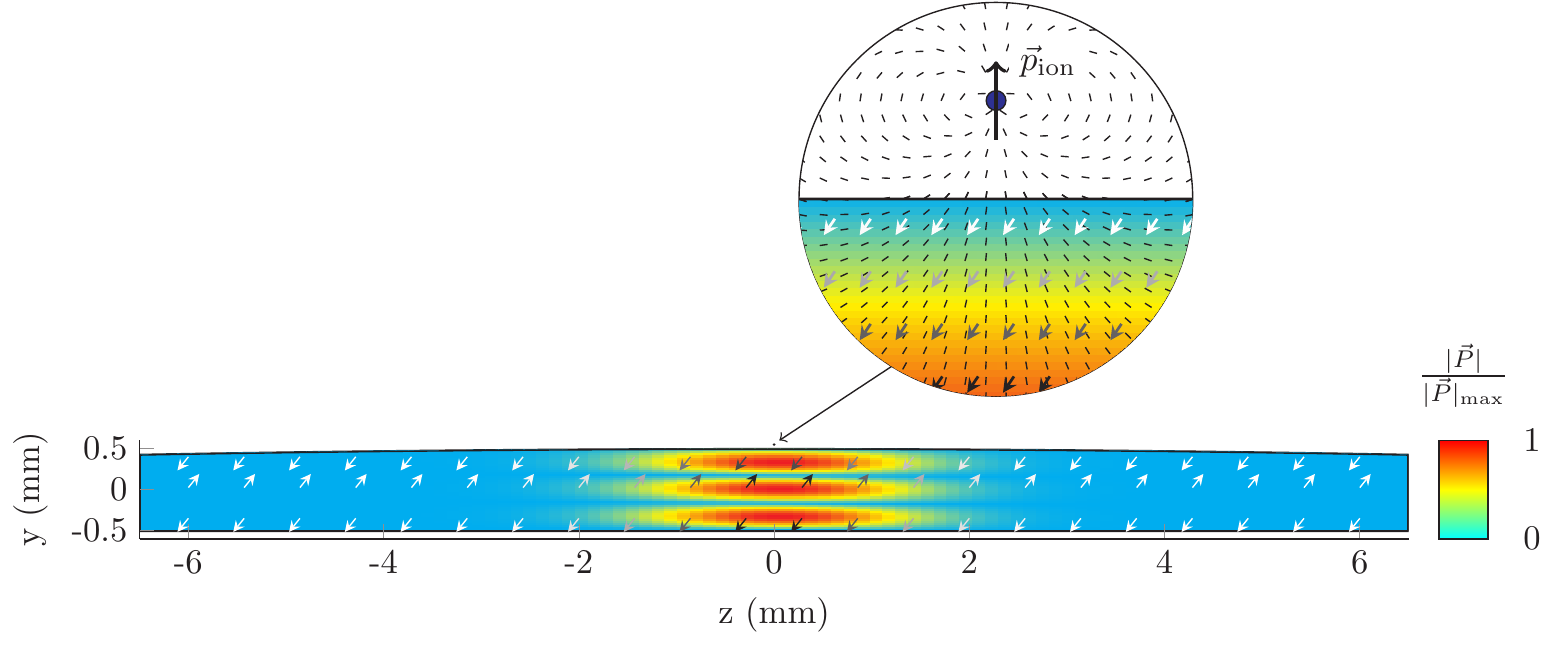}
			\caption{Piezo-electrically induced polarization density $\vP$ for third overtone acoustic mode (\Eref{eq:bvamode} with $n=3$). Magnitude (relative) is shown by the color plot. Direction is shown by the unit-vector arrows (arrow brightness indicate field strength). Inset: mode overlap between the electric-dipole field due to a fixed dipole $\pion$ at the ion position, which is associated with its motion along $\hat{y}$, and the quartz resonator polarization density $\vP$. Ion is assumed to be trapped $50\ \mu\un{m}$ above the resonator surface. The integral over the dipole-dipole interaction between $\pion$ and $\vP$ (\Eref{eq:piezo_as_dipole_dipole}) yields a coupling strength $g/2\pi\le 1~\un{Hz}$ (see appendix~\ref{apx:quart2ion}).}\label{fig:bvaionoverlap}
	\end{figure*}

This bound, however, cannot be saturated when using the actual integrand in~\Eref{eq:piezooverlap}. To see this, recall~\Eref{eq:piezo_as_dipole_dipole} where $g$ is expressed as an integral over the dipole-dipole interaction between the dipole defined by the ion motion, $\pion$, and the piezo-electrically induced polarization density $\vP$. Figure~\eref{fig:bvaionoverlap} illustrates the structure of $\vec{P}$. Naturally its magnitude follows that of the acoustic mode, having a Gaussian radial profile and forming a standing wave along $\hat{y}$-axis. The polarization direction of each standing-wave anti-node is approximately constant and opposite to that of its neighboring anti-nodes. Based on this structure, we can refine our upper bound for $g$ using

\begin{equation}
g\leq \frac{2\abs{\vec{p}_{\myr{ion}}}\abs{\vec{P}_{\myr{max}}}}{4\pi\hbar\bar{\epsilon}}\int_{V}\frac{d^3r}{r^3}\approx 3.2 \frac{\abs{\vec{p}_{\myr{ion}}}\abs{\vec{P}_{\myr{max}}}}{4\pi\hbar\bar{\epsilon}},
\end{equation}
where we used the fact that the interaction energy between two dipoles obtains a maximum when they are aligned with the vector $\vec{r}$ connecting them. For the mode configuration in~\Fref{fig:bvaionoverlap}, we get $g\le 2\pi\times 1.7~\un{Hz}$. 
This bound is confirmed in~\Aref{apx:quart2ion}, where we numerically calculate the coupling strengths for various ion motion axes according to~\Eref{eq:piezo_coupling_overlap_integral} and get $g/2\pi$ in the range of $0.49~\un{Hz}-1.46~\un{Hz}$. 

\begin{figure}[!h]\begin{center}
			\includegraphics{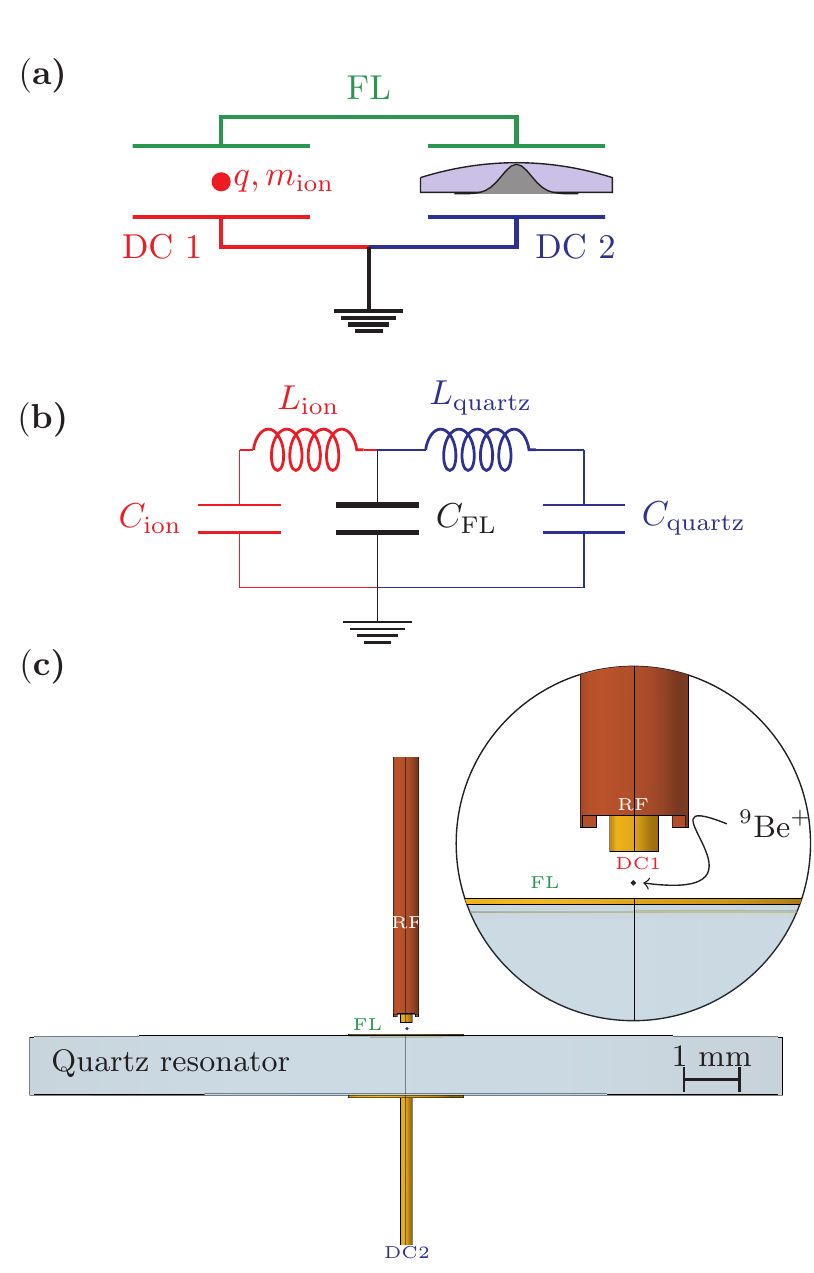}
		\caption{Coupling an ion to a quartz resonator mediated by a shunt capacitor. \capfont{a} The ion is trapped between two endcap electrodes forming a capacitor between FL and DC1. Ion motion generates image currents in the wires connecting the trap endcap DC1 and the quartz shunt capacitor (formed between FL and DC2), that in turn generate an oscillating electric field at the quartz resonator, actuating its acoustic modes through piezoelectricity \capfont{b} BVD equivalent circuit of the two coupled system. The capacitance $C_{\myr{FL}}$ is the total capacitance between FL and ground. \capfont{c} A Paul trap design minimizing $C_{\myr{FL}}$ for maximal coupling of a \Be ion to the quartz resonator. The trap is formed from a circular inner dc electrode (DC 1), surrounded by an outer cylindrical shell rf electrode (RF). Two disks of $1~\un{mm}$ radius placed at the top (FL) and bottom (DC 2) of the quartz resonator form the quartz shunt capacitance. Ideally, the top plate should be kept floating (FL) or connected to ground by a large ($>\un{G}\Omega$) resistor. The trap drive circuity that connects to the RF electrode and the RF ground connection between DC 1 and DC 2 is omitted.}\label{fig:ioncapbva}
		\end{center}
	\end{figure}

In order to increase the coupling strength, one could reshape the dipole field associated with the trapped ion, to better match the acoustic mode polarization density. A simple and practical way to do this is to use a capacitor to mediate the electric fields between the ion and the quartz resonator (see~\cite{Heinzen:1990zz}, appendix C), as in~\Fref{fig:ioncapbva}. Here the ion motion generates image currents on the trap electrodes that generate a time-varying, but uniform, electric field near the center of the crystal. 
	
The coupling $g$ can be calculated directly as done in~\ref{subsec:ionquartzcapacitive}. However, since the BVD equivalent capacitance $C_{\myr{quartz}}$ and inductance $L_{\myr{quartz}}$ of the quartz resonator have been measured for various acoustic modes, we present here a simpler analysis based on the BVD equivalent circuit of both the ion and the quartz resonator, shown in \Fref{fig:ioncapbva}\capfont{b}. Rewriting \Eref{eq:capacitivecoupling} for this case,

\begin{equation}
g=\frac{\omega_0}{2}\frac{\sqrt{C_{\myr{ion}}C_{\myr{quartz}}}}{C_{\myr{trap}}+C_{\myr{shunt}}},\label{eq:bvacapacitivecoupling}
\end{equation}
where we used the fact that the trap and shunt capacitance are much larger than the mechanical equivalent capacitances $C_{\myr{ion}}$ and $C_{\myr{quartz}}$. In fact, $C_{\myr{ion}}<0.2~\un{aF}$ (see~\Eref{eq:cionlion}) and typical values for $C_{\myr{quartz}}$ are in the $1-200~\un{aF}$ range~\cite{Goryachev:2011tb,Galliou:gVGkGc_h}. Therefore, it is imperative that the sum of the trap and shunt capacitance $C_{\myr{total}}\equiv C_{\myr{trap}}+C_{\myr{shunt}}$ are kept to a minimum. On the other hand, the quartz capacitor has to be large enough so as to have considerable overlap with the quartz acoustic mode. Since the mode radius is on the order of $\sigma\sim 1~\un{mm}$, the capacitor plate area should have a comparable radius, leading to $C_{\myr{shunt}}\sim 0.13~\un{pF}$, given the dielectric constant of these crystals $\epsilon=4\times 10^{-11}~\un{Fm}^{-1}$. The trap capacitance, therefore, should be comparable or lower than that value. \Fref{fig:ioncapbva}c shows an ion trap design where these low capacitances can be realized. The crux of the design is that instead of forming a trap capacitor separate from the quartz resonator capacitor and connecting them with wires, the top capacitor plate of the BVA also serves as the trap bottom dc plate. This arrangement is therefore able to minimize the effect of additional stray capacitances. Using an electrostatic simulation, we estimate $C_{\myr{total}}=0.18~\un{pF}$. 
	
The capacitor reshaping of the ion electric field indeed improves the coupling to $10-20~\myr{Hz}$ for known parameters of $C_{\myr{quartz}}$. With $N$ ions we get $gQ/2\pi\sim \sqrt{N}\times 10^{10}~\un{Hz}$, requiring a Wigner crystal of more than $100$ ions in order to satisfy the strong coupling regime constraint at $4~\un{K}$. Maintaining such a crystal in the trap might not be trivial due to the anharmonicities and finite size of the trap. In \ref{apx:quart2ion}, we show that the coupling dependence on different device parameters and mode overtone number does not allow for substantial increases in $g$. It has been shown that high overtone modes, e.g. $n=65$, can exhibit quality factors of almost $Q\sim 10^{10}$ \cite{Galliou:2013br}. That high-Q is counteracted by the $n^{-0.5}$ dependence of $g$ in the mode number (see appendix~\ref{apx:quart2ion}).

Nonetheless, it is worth noting the outstanding properties of such a device. The mechanical mode which is resonantly coupled to the ion motion can potentially be cooled to near its ground state by laser cooling the ion. Since laser cooling can be done much faster than the coupling rate, the quartz cooling rate is close to $2g/2\pi$. Thermal heating rate is $(1-e^{-1})n_{\myr{thermal}}\tau_{\myr{thermal}}^{-1}\approx (1-e^{-1})k_B T/(\hbar Q)$ (see \Sref{sec:couplinginthestrongcouplingregime}). The steady state number of quanta of the quartz acoustic mode would therefore be
\begin{equation}
\bar{n}\approx \pi(1-e^{-1}) \frac{k_BT}{\hbar g Q}.
\end{equation}
If operated at $4~\un{K}$, the $5-15~\un{MHz}$ mechanical modes of the quartz resonator could be cooled to $\bar{n}\sim 16~\myr{quanta}$ by laser cooling the coupled ion. Starting at dilution-refrigerator temperatures ($<50~\un{mK}$) would result in $\bar{n}\sim 0.2~\myr{quanta}$. The mechanical coherence times $\tau_{\myr{coh}}=\hbar Q/k_BT$ could reach $\sim 2~\un{ms}$ in a $4~\un{K}$ environment and up to $150~\un{ms}$ in a $50~\un{mK}$ environment. Due to its very large mode mass ($1-10~\un{mg}$), such a device, if placed in a superposition state of motion, could be used to restrict certain decoherence theories of massive objects (see~\Sref{sec:conclusion}). 

\section{Practical considerations for coupling an electron to a superconducting resonator}\label{sec:electron}
In \Sref{sec:supercondresonator}, we concluded that based on its small mass, the electron is potentially the most favorable candidate for a strongly coupled hybrid system composed of a charged particle and a superconducting resonator. Coupling strengths on the order of $0.1-1~\un{MHz}$ can be expected, requiring a very moderate quality factor of $Q\ge 10^4$ for the electrical resonator, at dilution-refrigerator temperatures. To estimate electron motion decoherence we extrapolate measured motional heating rates for trapped ions to an electron with a secular oscillation frequency of $1~\un{GHz}$ and get a heating rate of $\dot{n}\sim 100~\un{quanta}\cdot\un{s}^{-1}$, well below the coupling rate (see~\ref{apx:electronmotionheating}). 

The idea of using trapped electrons as part of a hybrid quantum system was first suggested for Penning traps~\cite{Heinzen:1990zz,Ciaramicoli:2003de}. To that end, novel planar Penning traps have been developed and demonstrated~\cite{Galve:2006eo,Galve:2007hp}. Moreover, electrons were trapped with cryogenic planar Penning traps~\cite{2008EPJD...50...97B}. Although single electrons have already been detected in three-dimensional Penning traps by driving their motion~\cite{Wineland:1973gfa,Brown:1986hc}, the anharmonicity of planar traps makes single electron detection challenging. An optimization of the design of the planar trap electrodes~\cite{Goldman:2010ky} led to the detection of one or two electrons~\cite{Goldman:2011wk}. The outlook for planar Penning traps is discussed elsewhere \cite{Marzoli:2009ce,2011EPJD...63....9B,Goldman:2011wk}.

Recently, an ensemble of $\sim 10^5$ electrons trapped on superfluid Helium with normal mode frequencies in the tens of GHz range, were non-resonantly coupled to a superconducting resonator at $\sim 5~\un{GHz}$~\cite{Yang:2016gw}. Measuring dispersive shifts in the resonator frequency in the presence of the electrons, the authors could deduce a coupling strength of $\sim 1~\un{MHz}$ per electron. Further studies of that technology could determine if the single electron regime can be achieved, establishing a new and interesting route for quantum information processing with electrons, as proposed in~\cite{Platzman:1999ce,Dahm:2002dq,Dykman:2003fl,Lyon:2006fs,Schuster:2010jn}.

The potential advantages and prospects of using rf Paul traps for electron-based quantum information processing were suggested and analyzed~\cite{Daniilidis:2013kw}. Clearly, since a Paul trap does not involve the strong magnetic fields required in a Penning trap, it naturally avoids exceeding the typical critical magnetic fields of superconducting circuitry. Strontium ions, for example, have been trapped with a superconducting Niobium planar chip trap~\cite{Wang:2010fa}. Two-dimensional trapping of electrons with rf fields was recently demonstrated, resulting in guiding electrons along a given trajectory~\cite{Hoffrogge:2011um}. To date, however, electrons have been almost exclusively trapped in three-dimensional Penning traps, with the exception of \cite{Walz:1995fm}. There, a macroscopic combined Penning and Paul trap was used to simultaneously trap tens of ions and electrons.

In~\cite{Daniilidis:2013kw}, a ring Paul trap design for electrons is analyzed, where a parametric coupling scheme is suggested, based on geometric nonlinearities of the potential. The coupling rates and decoherence rates reported here are compatible with those in that paper. The trap volume used in~\cite{Daniilidis:2013kw} was relatively small [$5\ \mu\un{m}\times \pi\times (15\ \mu\un{m})^2$] with a trap depth of $1~\un{meV}$, placing the electron $5\ \mu\un{m}$ away from the nearest electrode, rendering a strong coupling of $g= 2\pi\times 1.1~\un{MHz}$.

Here, we analyze the experimental conditions of two trap geometries, aimed at achieving the strong coupling regime, for a larger trapping volume and a deeper trap. As will be apparent in what follows, the design of these traps involves a delicate interplay between the trap stability and depth, its ability to maintain superconductivity, the energy range of the electron source, and the strong coupling requirement. In broad strokes, it is easier to build a big trap that is stable and deep so that currently available electron sources can be used. Large trap dimensions, however, would prevent satisfying the coupling criteria in~\Eref{eq:gQ}. On the other hand, a small trap is optimal for strong coupling but it can only support a shallow trapping potential and therefore requires a low energy electron source to ensure trapping. Because these problems are intertwined, our presentation includes a discussion of each of these aspects, as well as their compatibility.

\subsection{Stable trapping of electrons}\label{subsec:stabletrapping}
A Paul trap~\cite{Paul1990} is formed when a time-varying voltage $V_{\myr{rf}} \cos(\Omega_{\myr{rf}}t)$ is applied to an electrode arrangement that gives a quadratic spatial dependence for the electric potential in the neighborhood of its electric field null point. For simplicity, we assume cylindrical symmetry and write the time varying potential in terms of the standard $(\rho,z)$ cylindrical coordinates,

\begin{eqnarray}
\phi&=&qV_{\myr{rf}} \cos(\Omega_{\myr{rf}}t)\Phi(\rho,z),\\ \nonumber
\Phi(\rho,z)&=&\beta\frac{\rho^2-2z^2}{d^2},\quad \myr{ for}\ \rho,z\ll d,
\end{eqnarray}
where $q$ is the electron charge, $\beta$ is a unit-less geometry prefactor ($\beta=1$ for an ideal quadrupole) and $d$ is the trap electrodes length scale (e.g. distance from the trap center to the nearest point of an electrode surface). The time varying field generates a confining potential provided that the Mathieu criterion for stability is satisfied~\cite{Paul1990}:
\begin{equation}
q_{\myr{mathieu}}\equiv \frac{8\beta q V_{\myr{rf}}}{md^2\Omega^2_{\myr{rf}}}<1\label{eq:mathieustability}.
\end{equation}
The confinement can then be described, to lowest order, by a time-independent pseudo-potential:
\begin{equation}
\phi_{\myr{pseudo}}=\frac{q^2V_{\myr{rf}}^2}{4m\Omega_{\myr{rf}}^2}\abs{\nabla\Phi}^2\label{eq:phi_pseudo},\\
\end{equation}
where $m$ is the electron mass. It follows that the pseudo-potential trap depth can be expressed as $D=q V_{\myr{rf}} q_{\myr{mathieu}}/ \zeta$, where $\zeta$ is a unit-less factor dependent only on the trap geometry. For a perfect quadrupole trap $D=q V_{\myr{rf}} q_{\myr{mathieu}}/6$, whereas, for example, for a planar ``five-wire" surface electrode trap~\cite{Amini:2008tv}, $D=q V_{\myr{rf}} q_{\myr{mathieu}}/404$.

The first constraint we consider is trap stability (\Eref{eq:mathieustability}). Since the electron mass is small compared to ions, either the trap voltage should be lowered or the trap scale $d$ and/or frequency $\Omega_{\myr{rf}}$ should be increased, as compared to ion traps, to maintain stability. Lowering the voltage would reduce the trap depth and increasing $d$ would diminish the coupling strength. Therefore it appears to be advantageous to increase the trap frequency to the gigahertz regime. 

The second parameter we consider is trap depth. Naturally, it is easier to trap electrons in a deeper trap. For that purpose, increasing $V_{\myr{rf}}$ is beneficial. Other constraints, namely the need to maintain superconductivity in the trap electrodes and circuitry, limit the maximal rf voltage to a few tens of volts (see section~\Sref{sec:maintaining_superconductivity}). Thus far, the shallowest Penning trap that was able to maintain trapped electrons, had a trap depth of $D\sim 1~\un{eV}$, the electrons being loaded first into a $5~\un{eV}$ deep trap whose voltages were subsequently lowered to form the $1~\un{eV}$ trap~\cite{Goldman:2011wk}. We therefore will require the trap depth to be at least $D\sim 1~\un{eV}$.

\begin{figure}[!h]\begin{center}
			\includegraphics{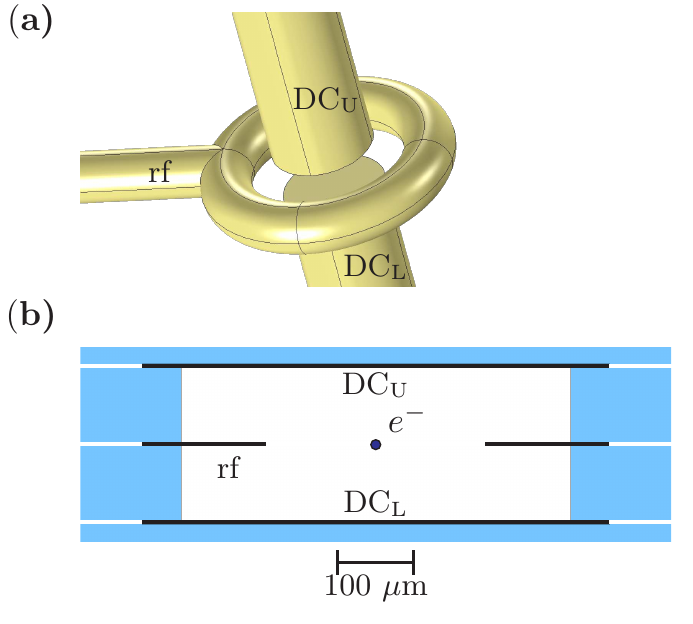}
			\caption{Two Paul trap designs for electron trapping. \capfont{a} An rf ring with $300\ \mu\un{m}$ inner diameter and $500\ \mu\un{m}$ outer diameter forms a quadrupole field at its center with respect to two dc endcaps. The flat-ended endcaps have a diameter of $200\ \mu\un{m}$ and are $100\ \mu\un{m}$ apart. \capfont{b} A two-dimensional cut through a stacked chip version of a. Blue region is a silicon substrate. The electron is trapped at the center of the middle rf ring electrode. The upper and lower endcap disks are $200\ \mu\un{m}$ apart. The center ring inner diameter is $240\ \mu\un{m}$ and the silicon-free region diameter is $500\ \mu\un{m}$. \Tref{tbl:electron3dtraps} summarizes the resulting trap parameters.}\label{fig:electron3dtraps}
		\end{center}
	\end{figure}

Figure \ref{fig:electron3dtraps} shows two different three-dimensional geometries of traps satisfying the above constraints. Table \ref{tbl:electron3dtraps} summarizes the resulting trap parameters. 
Figure \ref{fig:electron3dtraps}(a) describes a three-dimensional configuration of electrodes similar to~\cite{Bergquist:1985gg}. Here, the trap endcap to endcap distance is set to $d=100~\mu\un{m}$ in order to yield reasonable coupling while keeping a minimum distance of $50~\mu\un{m}$ between the ion and the nearest electrode to avoid large heating rates. The coupling also benefits from having no nearby dielectrics thereby minimizing the trap capacitance. The challenge in constructing such a trap, however, is the tolerance required for holding and aligning the electrodes. One way to solve this is shown in \Fref{fig:electron3dtraps}(b) where a trap is constructed from stacked chips, with lithographically patterned metal electrodes, pressed and aligned together~\cite{Rowej:2002wj,Britton:2010ds}. Since convenient wafer thickness is $\ge 100~\mu\un{m}$, the coupling is lowered since $d=200~\mu\un{m}$ and the trap capacitance increases due to the dielectrics involved. 
 
\begin{table}
	\begin{ruledtabular}
	\begin{tabular}{lll}
		Parameter                & Trap design in \Fref{fig:electron3dtraps}a & Trap design in \Fref{fig:electron3dtraps}b \\ \hline
		$V_{\myr{rf}}$           & $50~\un{V}$                                & $50~\un{V}$                                \\
		$I_{\myr{rf}}$           & $42~\un{mA}$                               & $243~\un{mA}$                              \\
		$\Omega_{\myr{rf}}/2\pi$ & $9~\un{GHz}$                               & $7.15~\un{GHz}$                            \\
		$\omega_{x,y}/2\pi$      & $0.6~\un{GHz}$                             & $0.75~\un{GHz}$                            \\
		$\omega_z/2\pi$          & $1.2~\un{GHz}$                             & $1.5~\un{GHz}$                             \\
		$D$                      & $1~\un{eV}$                                & $0.9~\un{eV}$                              \\
		$q_{\myr{mathieu}}$      & $0.4$                                      & $0.6$                                      \\
		$C_{\myr{trap}}$         & $15~\un{fF}$                               & $108~\un{fF}$                              \\
		$d$                      & $100\ \mu\un{m}$                           & $200\ \mu\un{m}$                           \\
		$g$                      & $2\pi\times 1.2~\un{MHz}$                  & $2\pi\times 203~\un{kHz}$                  \\ 
	\end{tabular}
	\end{ruledtabular}
	\caption{Trap parameters for the designs shown in \Fref{fig:electron3dtraps}. The pseudo-potential secular frequencies are $\omega_{x,y,z}$ where $x,y$ are in the plane of the rf ring and $z$ is perpendicular to it, $D$ is the trap depth, $C_{\myr{trap}}$ [see~\Eref{eq:ioncoil}] is the inherent total capacitance between the dc endcaps and $g$ is the electron-superconducting resonator coupling rate [see~\Eref{eq:ioncoil} as well as~\Fref{fig:trapAndDet} for circuit schematics]. With the above choices of $d$, the geometric parameter in \Eref{eq:ioncoil} is $\alpha\sim 1$ for both traps. Parameters are estimated using an electrostatic simulation (this is a reasonable approximation since in both traps the rf wavelength is $>10~\un{cm}$, i.e much larger than trap dimensions).The maximal rf current $I_{\myr{rf}}$ is estimated based on $I_{\myr{rf}}=\Omega_{\myr{rf}}C_{\myr{trap}}V_{\myr{rf}}$. Additional capacitance would result in higher current values.\label{tbl:electron3dtraps}}
\end{table}
 
\subsection{Maintaining superconductivity\label{sec:maintaining_superconductivity}}
An immediate concern with the above designs is that the relatively high rf currents involved will generate dissipation and potentially breakdown of the superconducting state of the trap electrodes. Usually, the electrodes of Paul traps form part of the capacitance $C$ of a parallel rf $LC$ resonator (e.g. in~\Fref{fig:trapAndDet} it would be the total capacitance between the two leads of $L_{\myr{rf}}$). We can estimate the on-resonance peak current $I_{\myr{max}}$ from the rf voltage amplitude $V_{\myr{rf}}$ using $\frac{1}{2}LI_{\myr{max}}^2=\frac{1}{2}CV_{\myr{rf}}^2$. We find $I_{\myr{max}}$ in the range of $200-400~\un{mA}$ for the conditions described below.

For simplicity, we restrict our analysis to thin film wires on chip, where an analytic treatment is available. The critical current, $I_c$, above which a thin film wire is no longer superconducting is 
\begin{equation} 
I_c=\frac{\Lambda\sqrt{wb}}{0.74}J_c,\label{eq:criticalcurrent}
\end{equation}
where $b$ is the film thickness, $w$ is its width, $\Lambda$ is the London penetration depth of the superconducting material and $J_c$ is its critical current density~\cite{Duzer1998}.

Of the two commonly used materials for superconducting devices, namely aluminum (Al) and niobium (Nb), aluminum is disadvantageous due to its lower values for $J_c$ and $\Lambda$ and since it requires operating in dilution refrigerators in order to superconduct (critical temperature $T_c=1.2~\un{K}$). For example, a $100~\un{nm}\times 10\ \mu\un{m}$ aluminum wire has a critical current of $I_c=11.3\ $mA. A niobium wire with the same dimensions would have a critical current of $I_c=221\ $mA and would be superconducting even at $4~\un{K}$ ($T_c=9.2~\un{K}$). 

To maintain superconductivity in the chip-based design in \Fref{fig:electron3dtraps}(b) with niobium films, we require thicknesses and widths that satisfy $bw>16\mu\un{m}^2$. Here, the features of the narrowest electrode or wire would serve as the bottleneck determining the critical current for the entire circuit. For example a $50\ \mu$m$\ \times\ 500\ $nm film cross section would be convenient to fabricate and would render $I_c=1.105\ $A. These numbers are compatible with those measured in a superconducting niobium trap for strontium ions~\cite{Wang:2010fa}.

Equation \eref{eq:criticalcurrent} actually constrains the dc critical current through a wire; however, the rf critical current for a superconducting resonator has similar values~\cite{Chin:1992fg}, at least for the case of a half-wavelength stripline resonator. Whether or not a similar result holds for a lumped element resonator where the current distribution is significantly different has yet to be demonstrated.  

\subsection{Low energy electron source}\label{sec:lowenergyelectronsource}
In principle, one method to load electrons into the trap would be to target the trapping volume with slow electrons and capture them by turning the trap on when they reach the trap center. In this case, the challenge lies in the fast electronics required. A slow electron source could be, for example, an ultra-cold GaAs photocathode~\cite{Weigel:2003wq,Orlov:2004df}, which has demonstrated beams with less than $1~\un{eV}$ average energy and less than $50~\un{meV}$ energy spread~\cite{Karkare:2015hh}. Such slow $0.1-1~\un{eV}$ electrons traversing a trap with a typical length of $100-200\ \mu\un{m}$, requires turning the trap on faster than a $0.1-1~\un{ns}$. In~\Sref{subsec:electricalcircuitry}, however,  we show that the trap resonator quality factor should exceed $10^4$ in order to comply with the typical cooling power of a cryogenic refrigerator. This would realistically limit the switching time of such a trap to the microsecond regime. 

One could mitigate this problem by constructing even slower electron sources. For example, using electron tunneling from bound states on the surface of liquid helium~\cite{Saville:1993jj} could potentially generate $<1~\un{meV}$ electrons, thereby relaxing the trap switching time constraint. The analysis of such a potentially novel source is beyond the scope of this paper.

A second type of electron source, which is commonly used in Penning traps, is based on secondary electrons~\cite{Walls:1973fh,RSVanDyck:1977gd}. For example, in~\cite{Goldman:2011wk}, a sharp tungsten tip was used to field-emit high energy ($\ge 200~\un{eV}$) electrons that collided with the trap surfaces, liberating gas molecules. During this process, some of these molecules reach the trapping region where they have a probability of being ionized by the incoming fast electrons. The relatively slow ``secondary" electrons generated in the ionization process could then be trapped.

This approach seems to be effective with deep ($\ge 5~\un{eV}$) and large ($d=0.1-2~\un{cm}$) traps~\cite{Goldman:2011wk}. Trap depth $\Ud$ is defined as the maximum minus the minimum of the trap pseudo-potential within the trap volume. It is not obvious that this technique would be efficient enough for a $\Ud=1~\un{eV}$ trap with a typical scale of $100\ \mu\un{m}$. Thus, we also consider a refinement of the secondary electron technique that might be less violent to the trap electrodes, as well as increase the trapping probability.

Rather than directing the incoming beam of electrons at the trap electrodes, we consider focusing the beam into the center of the trapping region and away from any surfaces. As a source of secondary electron emitters, a cold charcoal adsorber containing helium might be used. Primarily used for pumping residual helium gas, a charcoal adsorber can be heated with a resistor in order to liberate some helium and increase its vapor pressure in the chamber~\cite{Pobell:356689}. Incoming electrons will ionize the helium gas and generate secondary electrons that could then be trapped. In~\Sref{subsec:electricalcircuitry} we show that in order to accommodate for the heat load generated by the trap, it should be operated at temperatures in the range of $1-4~\un{K}$ and not dilution temperatures. That would also leave enough cooling power to remove the heat generated by the charcoal heating resistor. We henceforth assume that the refrigerator is operated at $4~\un{K}$.

The total cross section for helium ionization is maximal when the incoming electrons have a kinetic energy of $E_p\sim 120~\un{eV}$~\cite{USDepartmentofCommerce:ZgyY1S6W}. Here, however, we are interested in maximizing the cross-section for generating low energy secondary electrons rather than the total ionization cross section. In fact, since the threshold ionization for helium is $\sim 24.58~\un{eV}$, it is not surprising that the low energy cross-section peaks at $E_p\sim 30~\un{eV}$~\cite{Grissom:1972bm,Shyn:1979br}. The incoming electron energy should therefore be set to around $30~\un{eV}$, resulting in an optimal cross section of $\sigma_{\myr{ion}}\sim 0.05\ {\angstrom}^2$ for secondary electrons with energy below $1~\un{eV}$~\cite{Grissom:1972bm}. The resulting ionizing rate of helium atoms within the trapping volume is
\begin{equation}\label{eq:gammaion}
\Gamma_{\myr{ion}}\simeq \frac{J\pi r_0^2}{q_e} n_{\myr{He}} l \sigma_{\myr{ion}},
\end{equation}
where $J$ is the incoming current density of electrons, $q_e$ is the electron charge, $r_0$ is the incoming electron beam radius, $l$ is the radius of the spherical trapping volume and $n_{\myr{He}}$ is the vapor density of helium atoms. We restrict the discussion to secondary electron generation due to interaction of helium with the primary incoming electron beam. Additional ionization events due to, for example, elastically scattered electrons, could only increase $\Gamma_{\myr{ion}}$. In the presence of the rf trap, the incoming electrons energy $E_p$ will be spread by less than $\pm 15~\un{eV}$ around $30~\un{eV}$ as shown in~\Aref{apx:eecrosssection}. This, in turn, could reduce the average value of $\sigma_{\myr{ion}}$ by $< 18\%$ to $\sigma_{\myr{ion}}> 0.041~\angstrom^2$ (see~\cite{Grissom:1972bm}). Equation \eref{eq:gammaion} can therefore be considered as an average estimate for $\Gamma_{\myr{ion}}$. In addition, trap rf voltage can deflect the incoming electrons, causing the average beam radius to expand to $r_1=\xi r_0$. Since the rf trap voltages $V_{\myr{rf}}$ considered in this paper have the same order of magnitude as $E_p/q_e$ (see~\Tref{tbl:electron3dtraps}) $\xi\le 4$ as shown in~\Aref{apx:eecrosssection}. We can still use $r_0$ in~\Eref{eq:gammaion} since it depends on the total current of electrons traversing the trapping region. As long as $r_1<l$, electrons are not lost due to collisions with the trap walls and this total current should be preserved.

The steady state number of trapped electrons is determined by the ratio between the low-energy secondary electron generation rate $\Gamma_{\myr{ion}}$ and the total electron loss rate. Electrons that have already been trapped may collide with incoming electrons or with the surrounding helium atoms. The average energy of the electrons gradually increases due to these collisions (heating) until eventually it exceeds the trap depth and they are lost (boiling).

In \Aref{apx:eecrosssection}, we derive analytically an upper bound on the contribution to the heating rate due to collisions with incoming electrons. Briefly, since each collision is a Rutherford-type scattering problem, it cannot be attributed a finite cross section. Its geometric scale is therefore dictated by the incoming electron beam finite radius $r$ where $r_0\le r\le r_1$. Therefore, the average energy a single trapped electron gains in a single collision is $< q_e^2/(4\pi\epsilon_0 r_0)$. Since the rate of collisions is $J\pi r_0^2/q_e$ the resulting heating rate is $(dE/dt)|_e<  J r_0 q_e/(4\epsilon_0)$. This translates to an electron loss rate of
\begin{equation}\label{eq:gammae}
\Gamma_e=\frac{1}{\Ud}\left.\left(\frac{dE}{dt}\right)\right|_e<\frac{J r_0 q_e}{4\epsilon_0\Ud}.
\end{equation}

The contribution to the heating rate due to collisions with the helium gas is known as ``rf-heating". This follows from the helium atom playing the role of a hard immovable ball in the collision process, being much heavier than the electron. Therefore, when an electron collides with it, its instantaneous micro-motion kinetic energy before the collision transforms into the secular motion energy after the collision~\cite{1968AdAMP...3...53D,1969AdAMP...5..109D}. During the harmonic secular motion of the ion, kinetic energy is exchanged between rf and secular motion, the rf fraction being maximal farthest from the trap center and ideally zero at the center. Therefore, collisions that occur farther from the center will potentially transfer more energy into the secular motion. If the secular energy of the trapped electron prior to collision is $E_{\myr{in}}$, the energy gain after a single collision is $\le E_{\myr{in}}/2$, when averaging over the secular motion period. Assuming that the trapped electrons have a uniform energy distribution between $0$ and $\Ud$, the average energy gain per collision with a single helium atom is smaller than $\Ud/4$.  The rate of collisions in this case is $\sim \sigma_{\myr{elastic}} n_{He} \langle\abs{v}\rangle$ where $\sigma_{\myr{elastic}}\sim 6\ {\angstrom}^2$ is the electron-helium elastic cross section for low energy ($\le 2~\un{eV}$) electrons~\cite{Shigemura:2014bt} and $\langle \abs{v} \rangle \sim \tfrac{4\sqrt{2}}{3\pi}\sqrt{{\Ud}/{m_e}}$ is the average velocity of the trapped electrons, $m_e$ being the electron mass. The resulting heating rate is $(dE/dt)|_{He}<\sigma_{\myr{elastic}} n_{He} \langle \abs{v}\rangle\Ud/4$. We translate it to an electron loss rate of
\begin{equation}\label{eq:gammeHe}
\Gamma_{He}<\frac{1}{\Ud}\left.\left(\frac{dE}{dt}\right)\right|_{He}=\frac{\sigma_{\myr{elastic}}n_{He}}{3\pi}\sqrt{\frac{2U_{\myr{depth}}}{m_e}}.
\end{equation}

Combining equations \Eref{eq:gammaion}-\eref{eq:gammeHe}, the steady state number of electrons in the trap, $N_e$, is dictated by setting $dN_e/dt=0$ in the rate equation
\begin{equation}\label{eq:Herateequation}
 \frac{dN_e}{dt}=\Gamma_{\myr{ion}}-N_e(\Gamma_{\myr{e}}+\Gamma_{\myr{He}}).
\end{equation}

\begin{figure}[!hbtp]\begin{center}
		\includegraphics{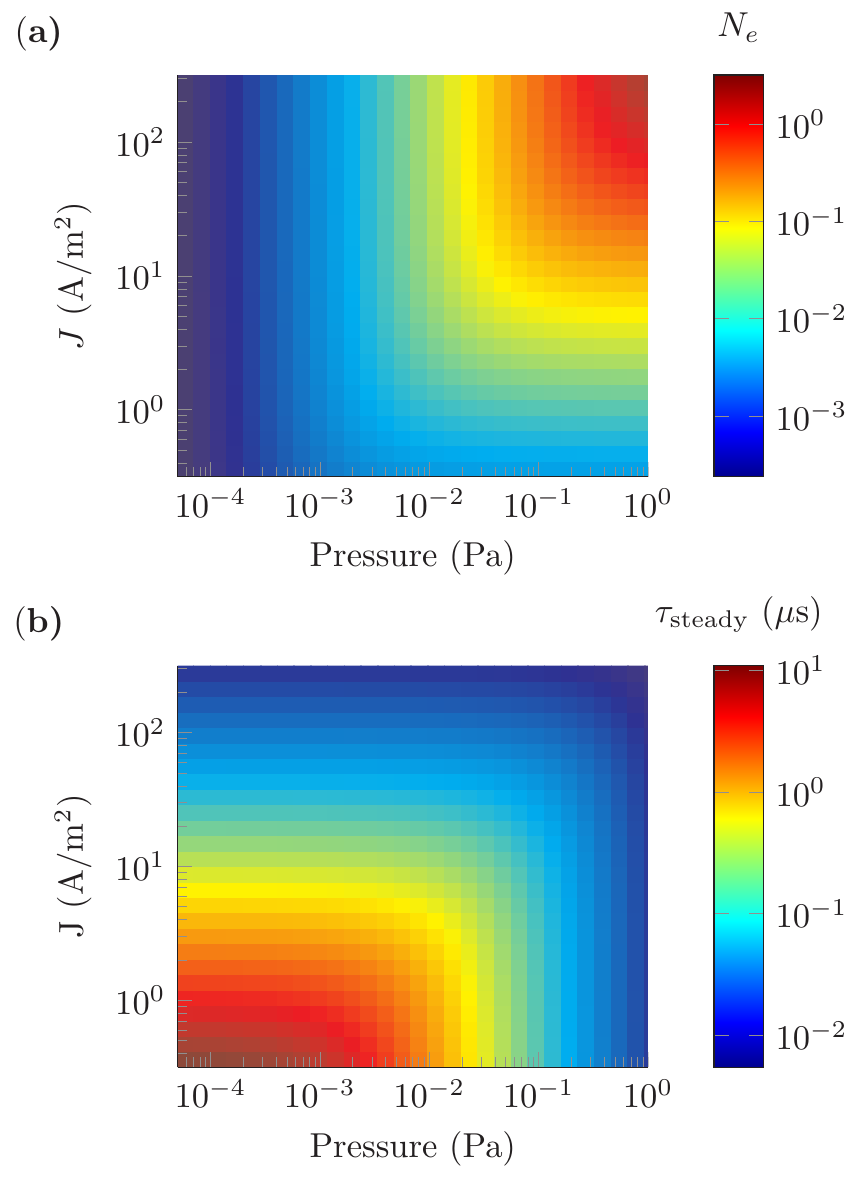}
		\caption{Effect of loading parameters. \capfont{a} Estimated steady-state number electrons $N_e$ in a $1~\un{eV}$-deep trap having a trapping volume of $\sim (95\ \mu\un{m})^3$, when the electron gun is on. Incoming electron beam radius is assumed to be $r_0=10\ \mu\un{m}$. \capfont{b} $1/e$-time to reach steady state number of electrons.}\label{fig:steadystateNe}
	\end{center}
\end{figure}	

For trapping, we require the steady state number of electrons $N_e$ be greater than a threshold value $N_{\myr{threshold}}$, as we discuss below. This can always be satisfied if the current density $J$ and the density of helium $n_{\myr{He}}$ are large enough [\Eref{eq:Herateequation}]. To see this quantitatively, in \Fref{fig:steadystateNe}\capfont{a}, we plot the number of steady state electrons for different current densities and helium pressure values. 

The value for $N_{\myr{threshold}}$ depends on the cooling rate of the electron motion $\Gamma_{\myr{cool}}$. Without cooling, once the incoming electron source is turned off ($J\to 0$), any trapped electrons would rapidly boil out of the trap due to collisions with the helium background gas. Indeed, the helium pressure can be decreased significantly to avoid this process by allowing the charcoal adsorber to cool to its $4~\un{K}$ surroundings. However, the time-scale for removing the helium is likely to be long  compared to $1/\Gamma_{He}$. The latter is inversely proportional to the helium pressure and, for example, equals $1.3~\mu\un{s}$ at a helium pressure of $10^{-2}~\un{Pa}$.

In the design we consider below, we assume the $z$ motion of the trapped electrons is strongly coupled to an LC-resonator to experience damping. In \Sref{sec:detection}, we show that a $\sim1~\un{GHz}$ LC resonator with a quality factor $Q_{\myr{det}}\sim 1000$ should suffice for single electron detection. Therefore, the LC resonator equilibrates with its $4~\un{K}$ surroundings at a $\sim 1~\un{MHz}$ rate, i.e. much faster than the coupling rate $g$ between the LC resonator and the electron motion. The resulting $z$-motion damping rate is dictated by the slower of the rates, $\Gamma_{\myr{cooling}}\sim g/2\pi \ge 100~\un{kHz}$. In order to cool the $x$ and $y$ motion, these modes could be parametrically coupled to the $z$ motion~\cite{Gorman:2014ba} as discussed in~\Sref{sec:parameteric_coupling}. We will henceforth assume a similar damping rate for all axes. 

Once the incoming electron beam is turned off, the trapped-electron energy $E$ is dictated by the cooling rate and the helium collision-induced heating rate:
\begin{equation}\label{eq:afterloadingheatinrate}
\frac{dE}{dt}=-\Gamma_{\myr{cool}}E+\frac{1}{\pi}(\sigma_{\myr{elastic}}n_{\myr{He}})\sqrt{\frac{2E}{m_e}}E.
\end{equation}
For this equation to be correct, the initial energy of the electron must be below a value $E_{\myr{init}}$ determined by trap anharmonicity, which manifests as an amplitude-dependence of the resonant frequency. Since damping is based on resonant coupling to the LC resonator, large amplitude motion will not cool effectively. Based on \Sref{sec:detection}, we can estimate $E_{\myr{init}}\lesssim 0.3~\un{meV}$.

To achieve net cooling, the right hand side of \Eref{eq:afterloadingheatinrate} should be negative, i.e.,
\begin{equation}
E\le E_{\myr{capture}}\equiv \frac{m_e}{2}\left(\frac{\pi \Gamma_{\myr{cool}}}{\sigma_{\myr{elastic}}n_{\myr{He}}}\right)^2.
\end{equation}
Therefore, if the electron $z$-motion satisfies $E<E_{\myr{thresh}}\equiv\min\left(E_{\myr{capture}},E_{\myr{init}}\right)$, it will remain trapped. 
For helium pressures below $0.027~\un{Pa}$, $E_\myr{init}$ is the smaller of the two and determines $E_\myr{thresh}=0.3~\un{meV}$. For a pressure $P$ greater than that, $E_\myr{thresh}=E_\myr{capture}=0.3~\un{meV}\times (0.027~\un{Pa}/P)$.

Equation~\eref{eq:afterloadingheatinrate} was based on the assumption that excess micromotion can be neglected. Excess micromotion occurs when the ion experiences rf fields even at its equilibrium position that is usually shifted from the rf-null due to stray fields. This would lead to a constant heating term in~\Eref{eq:afterloadingheatinrate}, thereby limiting both $E_{\myr{capture}}$ as well as the steady-state energy. Using dc compensation fields, the ion position can be adjusted back to the rf null. We require the heating rate due to excess micromotion to be much lower than the heating rate for electrons with $E_{\myr{capture}}$ energy. If the ion is at a position $x$ away from the rf null, this constraint can be written as $m_ev_\myr{mm}^2(x)\ll E_{\myr{capture}}$ where $v_\myr{mm}(x)$ is the micromotion velocity amplitude at $x$. For a $1~\un{GHz}$ trap and $E_{\myr{capture}}=0.3~\un{meV}$ this constrains $x\ll 1~\mu\un{m}$.

\begin{figure}
	\includegraphics{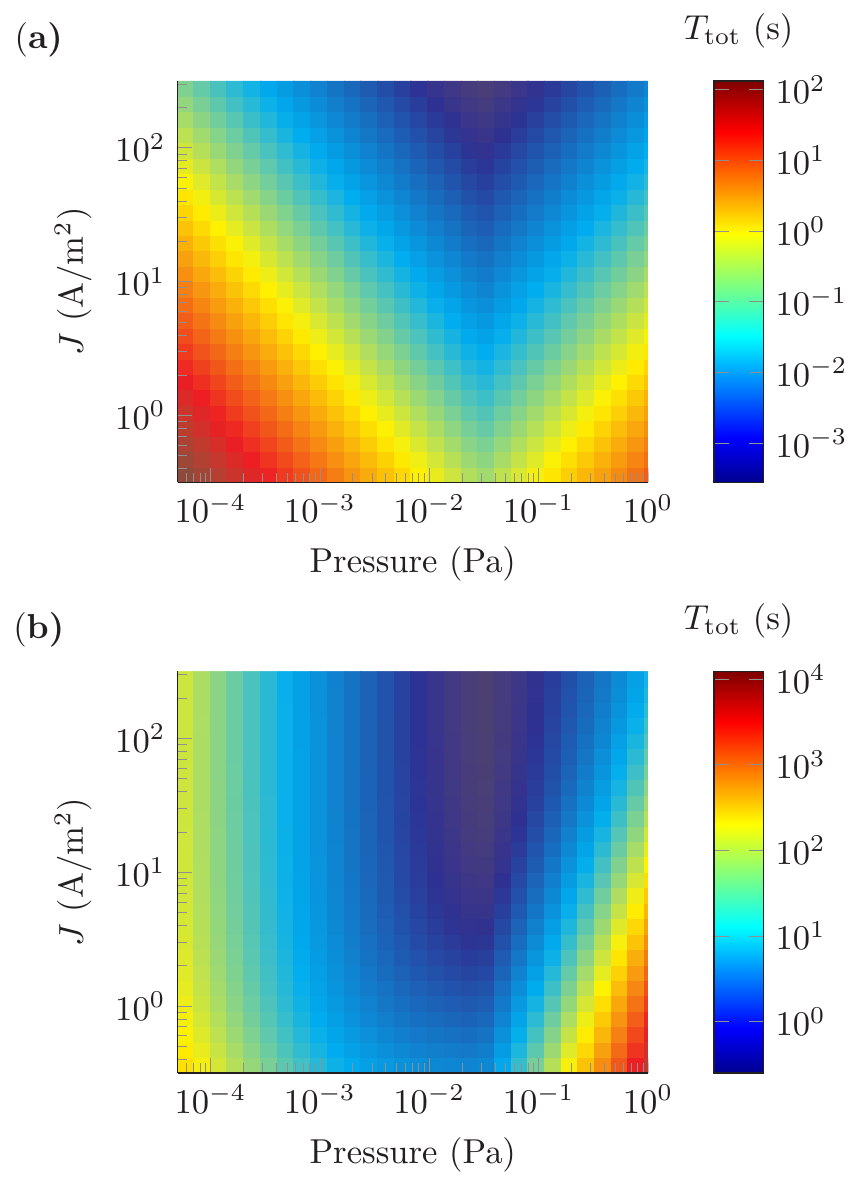}
	\caption{Estimated average total time $T_{\myr{tot}}$ for trapping and detecting a single electron, based on the same parameters used for~\Fref{fig:steadystateNe}. The incoming electron beam gun is operated in pulse mode, the duration of each pulse [\Fref{fig:steadystateNe}\capfont{b}] allows to reach a steady state number of electrons [\Fref{fig:steadystateNe}\capfont{a}]. This translates into a probability of trapping a single electron after a single pulse. The process must be repeated a number of times which is inversely proportional to that probability. After the electron loading pulse, a detection procedure needs to be applied for $T_{\myr{det}}$. \capfont{a} Assuming $T_\myr{det}=0$, i.e. negligible. \capfont{b} Assuming $T_\myr{det}=10~\mu\un{s}$ based on the conservative end of detection-time estimates from~\Sref{sec:detection}.}\label{fig:timetotrapone}
\end{figure}	

From~\Frefs{fig:steadystateNe}\capfont{a} and \capfont{b} we can extract the time needed to trap a single electron. Within the parameters explored, the steady state number of trapped electrons $N_e$ is less than one and the threshold energy is $E_{\myr{thresh}}\sim 0.3~\un{meV}$ or smaller. Therefore, the loading process should be operated in pulsed mode, with $\sim (\Ud/E_{\myr{thresh}})/N_e$ pulses required on average to trap a single electron (provided that the electron  energy distribution is uniform between zero and $\Ud$). Combined with the $1/e$ time required to reach the steady state [\Fref{fig:steadystateNe}\capfont{b}], we extract the average total time required for trapping a single electron, shown in~\Fref{fig:timetotrapone}\capfont{a}. As long as $E_{\myr{thresh}}$ is not dominated by the helium pressure $P$, i.e. by $E_{\myr{capture}}$, increasing $P$ is beneficial since $N_e$ increases.  An optimal helium pressure of $\sim 0.027~\un{Pa}$ is reached, beyond which $E_{\myr{thresh}}=E_{\myr{capture}}\propto 1/P^2$. 

These estimates assume that once a single electron is trapped, it is immediately detected. Realistically, some sort of detection procedure needs to be applied in order to verify that indeed an electron is present. In~\Sref{sec:detection} we analyze a the detection scheme of~\cite{Wineland:1975kk}. We estimate that the time to detect a single electron $T_{\myr{det}}$ is in the $1-10~\mu\un{s}$ range. In~\Fref{fig:timetotrapone}\capfont{b} we plot the total time required to trap and detect a single electron for the more conservative estimate for $T_{\myr{det}}=10~\mu\un{s}$ . Based on the plot, working in the helium pressure range of $10^{-4}-10^{-1}~\un{Pa}$ and the current density range of $1-100~\un{A}/\un{m}^2$, the range of times we get is similar to that of Paul trap loading times for ions. 

The current density range in~\Frefs{fig:steadystateNe}-\ref{fig:timetotrapone} is chosen such that the total current of incoming electrons is in the nano-amps regime for a beam radius of $r_0=10\ \mu\un{m}$. The beam radius was chosen so that even after expansion to $r_1$ due to the trap rf fields it would avoid the trap walls. These parameters can be easily obtained with commercial electron sources. Smaller beam radii with the same total current would reduce the total time required to trap an electron even further. That would require a design of electron optics combined with either a commercial or home made cold field emission source, the details of which are beyond the scope of this paper. 

\subsection{Electrical circuitry}\label{subsec:electricalcircuitry}
Stable trapping requires applying large voltages and currents in a cryogenic environment, next to a sensitive detection resonator. This has implications on the refrigerator heat load and the circuit design of the trap.

Achieving a trap drive amplitude of $V_{\myr{rf}}=100~\un{V}$ at frequencies in the $7-9~\un{GHz}$ range requires resonating the trap capacitance $C_{\myr{rf}}$ with an inductor. The resulting dissipation rate would be $P_{\myr{dis}}=\Omega_\myr{rf}C_{\myr{rf}}V_{\myr{rf}}^2/Q$ where $Q$ is the rf resonator quality factor. With $C_{\myr{rf}}\le 150~\un{fF}$ (based on simulations of the traps in \Fref{fig:electron3dtraps}) and $Q$ in the $10^4-10^5$ range this implies $\le 0.2-2~\un{mW}$ of dissipated power for frequencies in the $7-9~\un{GHz}$ range. With the cooling power of a dilution refrigerator typically being in the $100\ \mu\un{W}-400\ \mu\un{W}$ range at $T=100~\un{mK}$, working at $4~\un{K}$ would be indicated where $2~\un{mW}$ of power dissipation is easily handled, even with a lower ($Q\sim 10^4$) quality factor. In fact, even $1-2~\un{K}$ cryostats with $\sim 60-200~\un{mW}$ of cooling power could suffice.

\begin{figure*}
		\includegraphics[width=0.8\textwidth]{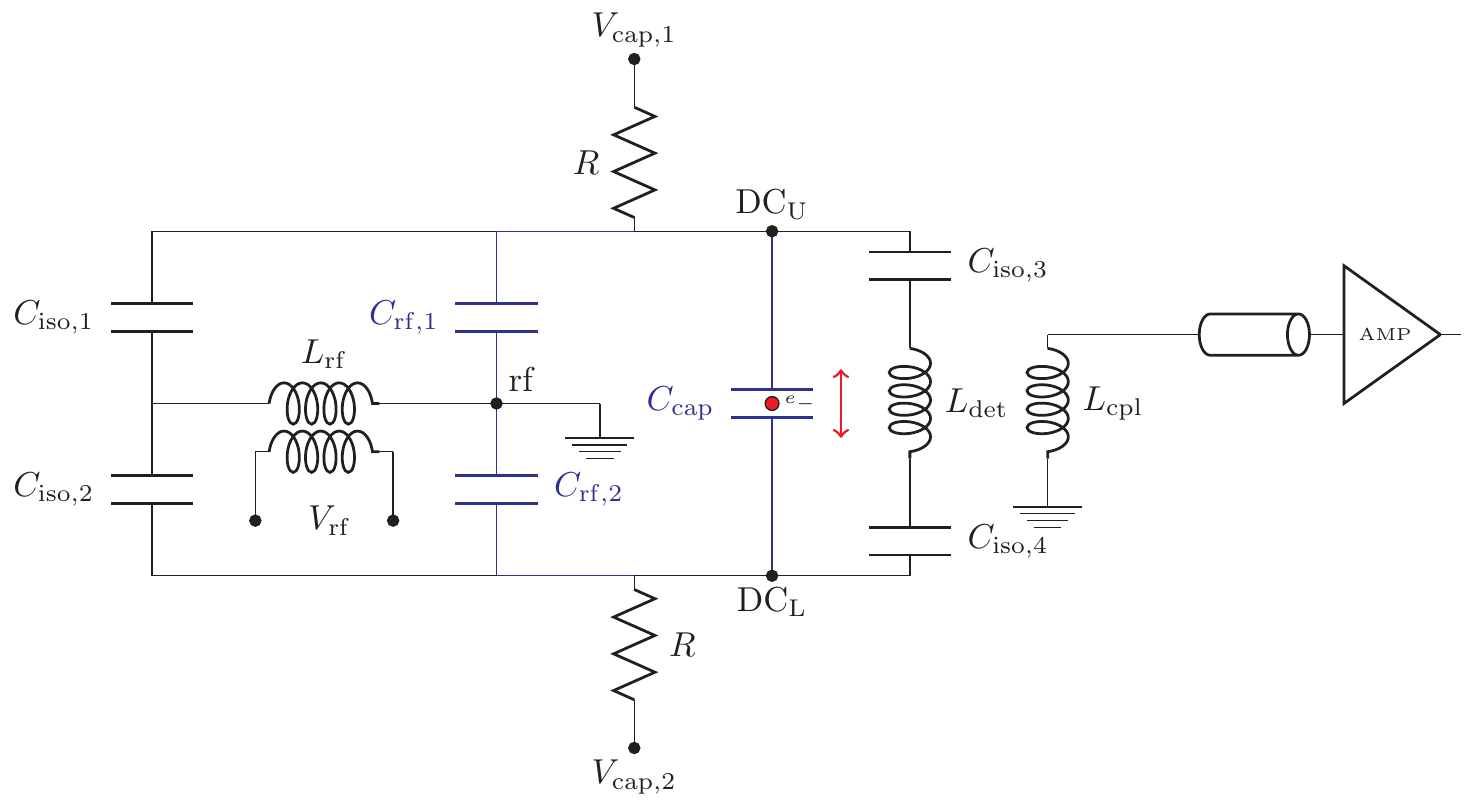}
		\caption{Trap and detection resonators schematic for the traps illustrated in \Fref{fig:electron3dtraps}. The electrodes $\myr{DC}_\myr{U}$, $\myr{DC}_\myr{L}$ and rf of \Fref{fig:electron3dtraps} are indicated here in the schematic. The trap capacitances are shown in blue where $C_{\myr{rf,1}}$ and $C_{\myr{rf,2}}$ are the capacitances between the center ring and each end-cap and $C_{\myr{cap}}$ is the end-cap to end-cap capacitance. For the trap in \Fref{fig:electron3dtraps}\capfont{a}, these equal to $21.3~\un{fF}, 21.3~\un{fF}, 4.6~\un{fF}$ correspondingly. For the trap in \Fref{fig:electron3dtraps}\capfont{b}, these equal to $146~\un{fF}, 146~\un{fF}, 35~\un{fF}$ correspondingly. The $L_{\myr{rf}}$ inductor forms a resonator with the total capacitance between its ends generating the quadrupole trapping field. The $L_{\myr{det}}$ inductor along with the capacitance shown forms the detection resonator that monitors the electron motion (double red arrow). The four isolation capacitors enable independent dc-biasing of the upper and lower end-caps ($V_{\myr{cap,j}},\ j=1,2$) with bias resistors $R\ge 10~\un{M}\Omega$ to avoid loading the detection circuit, assumed to have a quality factor of $\sim 1000$ [see~\Sref{sec:detection}]. The leftmost isolation capacitors $C_{\myr{iso,1}}$ and $C_{\myr{iso,2}}$ are chosen to equal $C_{\myr{rf,1}}=C_{\myr{rf,2}}$. The rightmost isolation capacitors $C_{\myr{iso,3}}$ and $C_{\myr{iso,4}}$ are chosen to be much greater than the total capacitance between  $\myr{DC}_\myr{L}$ and  $\myr{DC}_\myr{U}$, e.g. on the order of $1~\un{pF}$. The mutual inductance of $L_{\myr{det}}$ and $L_{\myr{cpl}}$ allows for the electron detection using an amplifier.}\label{fig:trapAndDet}
\end{figure*}

To understand the implications of the trap drive on the electron detection circuit, we model the traps in \Frefs{fig:electron3dtraps}\capfont{a}\&\capfont{b} with a lumped element circuit shown in \Fref{fig:trapAndDet}. Detecting the presence of electrons would be accomplished using a tank circuit technique~\cite{Wineland:1975kk,Brown:1986hc}. The electron thermal motion generates image currents that couple to the resonator formed from the trap capacitance and the inductor $L_{\myr{det}}$, chosen to be resonant with the $\sim 1~\un{GHz}$ secular motion. The trap is driven by a different resonator, formed from the ring-to-end-caps capacitance and another inductor, $L_{\myr{rf}}$, chosen to resonate at the $7-9~\un{GHz}$ drive frequency. 

The possible cross talk between the drive and detection resonators could deteriorate their respective quality factors. If the trap is electrically symmetric, i.e. $C_{\myr{rf},1}=C_{\myr{rf},2}$ and $C_{\myr{iso},1}=C_{\myr{iso},2}$, the two circuits are essentially orthogonal. The detection circuit is connected to equi-potential points in the trap drive circuit and is therefore not influenced by the high currents flowing there. Moreover, due to the Wheatstone bridge topology, the detection circuit is not sensitive to the rf inductor $L_{\myr{rf}}$ and its coupling port. It is only influenced by the additional capacitances $C_{\myr{iso},j}$ for $j=1,2$ that add to the total trap capacitance. Similarly, the rf resonator is indifferent to the added impedance of the detection resonator. The impact of trap asymmetry on the quality factor of the two resonators can be estimated by:
\begin{subequations}
	\begin{eqnarray}
\frac{\Delta Q_{\myr{rf}}}{Q_{\myr{rf}}}&\sim& \frac{Q_{\myr{rf}}\omega_0}{Q_{\myr{det}}\Omega_{\myr{rf}}} \frac{C_{\myr{cap}}}{C_{iso,1}+C_{rf,1}+2C_{cap}}  \epsilon,\\
\frac{\Delta Q_{\myr{det}}}{Q_{\myr{det}}}&\sim& \frac{Q_{\myr{det}}\omega_0}{Q_{\myr{rf}}\Omega_{\myr{rf}}} \frac{C_{\myr{iso,1}}+C_{\myr{rf,1}}}{C_{\myr{iso,1}}+C_{\myr{rf,1}}+2C_{\myr{cap}}} \epsilon,\\
\epsilon&=& \frac{\abs{C_{rf,1}-C_{rf,2}}+\abs{C_{iso,1}-C_{iso,2}}}{C_{rf,1}+C_{iso,1}},
\end{eqnarray}
\end{subequations}
where $Q_{\myr{rf}}$ and $Q_{\myr{det}}$ are the rf and detection resonator quality factors respectively when the trap is completely symmetric, $\Delta Q_{\myr{rf}}$ and $\Delta Q_{\myr{det}}$ is their respective change due to asymmetry, $\omega_0\sim 2\pi\times 1~\un{GHz}$ is the secular frequency, $\Omega_{\myr{rf}}\sim 2\pi\times 7-9~\un{GHz}$ is the trap drive frequency and $\epsilon$ is the asymmetry parameter. Clearly, if $Q_{\myr{rf}}$ and $Q_{\myr{det}}$ are comparable, and the capacitances involved are on the same order of magnitude, then keeping $\epsilon$ below a few percent should suffice. 

\subsection{Non-linearity and detection of a single electron}\label{sec:detection}
One of the main concerns with detecting a single electron in Penning trap experiments is the trap anharmonicity~\cite{2008EPJD...50...97B,Marzoli:2009ce,Goldman:2010ky}. In these traps, the signal of a single electron has a few hertz linewidth due to damping resulting from its coupling to the detection circuit, whereas the effect of anharmonicity in these planar traps is to broaden the electron detection signal to $10~\un{kHz}-1~\un{MHz}$. However, in~\cite{Goldman:2010ky}, it was shown that by adding compensation electrodes and carefully adjusting their relative voltages, one could avoid the dominant anharmonic terms of the potential. Similarly, careful consideration for electrode shape and geometry allow for higher degree of harmonicity in three-dimensional traps~\cite{Beaty:1986ir,Beaty:1987ip}. 

\begin{figure}[!hbtp]\begin{center}
		\includegraphics[width=0.3\textwidth]{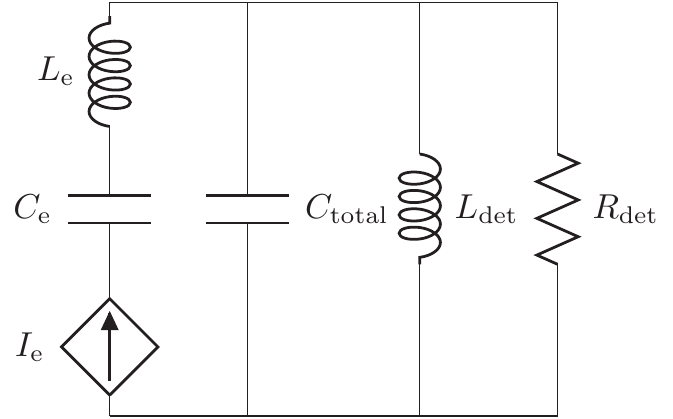}
		\caption{Simplified electron detection circuit, based on the circuit in~\Fref{fig:trapAndDet}. Here, $C_{\un{total}}$ is the total capacitance between the two ends of the detection inductor $L_{\myr{det}}$. The trapped electron is replaced by its electrical equivalent of a series LC resonator with inductance $L_\myr{e}$ and capacitance $C_\myr{e}$. Currents generated by ion motion are represented by $I_\myr{e}$. The coupling inductor $L_{\myr{cpl}}$ in~\Fref{fig:trapAndDet} transduces the input impedance of the amplifier to an effective resistance, which, combined with LC internal dissipation, are represented by an equivalent shunt resistor $R_{\myr{det}}$. }\label{fig:electronDetection}
		\end{center}
\end{figure}

In the designs considered here, the electron is strongly coupled to the detection circuit, giving a relatively broad signal linewidth which in turn relaxes the constraints on the trap harmonicity. By assuming a moderate quality factor for the detection circuit $Q_{\myr{det}}\sim 1000$, the detection circuit linewidth is on the order of $\sim 1~\un{MHz}$ and therefore larger than anharmonicity induced broadening of the electron signal as we show below. In order to reach the strong quantum regime, however, we required $Q_{\myr{det}}\gg 7000$ (see~\Tref{tbl:ioncoil}). However, with a tunable coupler~\cite{Yin:2013jv}, one could potentially tune the quality factor of the detection circuit to accommodate for both $Q$-factor regimes. Detailed analysis of such a coupler is beyond the scope of this paper. Therefore, in this section and in~\Sref{sec:parameteric_coupling} we use the lower $Q_{\myr{det}}\sim 1000$ value.

Figure \ref{fig:trapAndDet} shows the schematics of a typical tank detection circuit and \Fref{fig:electronDetection} shows a simplified equivalent circuit. The simplification follows first from replacing the trapped electron with its BVD equivalent network $L_e,C_e$ and a current source $I_e$ corresponding to the induced currents due to ion motion. Further simplification is achieved by replacing the entire network connected to the two ends of the detection inductor $L_{\myr{det}}$ with its total equivalent capacitance $C_{\myr{total}}$. This will define the tank circuit resonant frequency $\omega_0=1/\sqrt{L_{\myr{det}}C_{\myr{total}}}$ which we assume to be resonant with the electron trap frequency. Finally, the amplification network which couples to $L_{\det}$ via mutual inductance to the coupling inductor $L_{\myr{cpl}}$ is replaced by an equivalent resistor $R_{\myr{det}}$. The coupling inductor $L_{\myr{cpl}}$ transduces the input impedance of the amplifier, the real part of which presents an effective resistance $R_{\myr{ext}}$ in parallel with the internal resistance $R_{\myr{int}}$ of the LC tank circuit. The total resistance of the detection circuit is therefore $R_\myr{det}= R_{\myr{ext}}R_{\myr{int}}/(R_{\myr{ext}}+R_{\myr{int}})$. The width of the electron signal can be estimated to be $R_{\myr{det}}/L_{\myr{e}}\sim 2\pi\times 100\ $kHz, expressed in terms of the trap parameters: 

\begin{equation}
\frac{R_{\myr{det}}}{L_{\myr{e}}}=\frac{Q_{\myr{det}}q_e^2\alpha^2}{\omega_0 C_{\myr{total}} m_ed^2},
\end{equation}
where $d\sim 200\ \mu\un{m}$ is the end-cap to end-cap distance, $\omega_0=2\pi\times 1~\un{GHz}$ is the trap secular motion frequency and $C_{\myr{total}}\sim 180~\un{fF}$ for the trap in~\Fref{fig:electron3dtraps}\capfont{b}. The capacitance $C_{\myr{total}}$ is calculated by expressing it in terms of the other capacitances in~\Fref{fig:trapAndDet}:
\begin{equation}
C_{\myr{total}}=C_{\myr{cap}}+\frac{C_{\myr{rf,1}}C_{\myr{rf,2}}}{C_{\myr{rf,1}}+C_{\myr{rf,2}}}+\frac{C_{\myr{iso,1}}C_{\myr{iso,2}}}{C_{\myr{iso,1}}+C_{\myr{iso,2}}},
\end{equation}
assuming that $C_{\myr{iso},k}$ ($k=3,4$) are much larger than $C_{\myr{trap}}$. 

While $C_{\myr{rf,k}}$, $k=1,2$ and $C_{\myr{cap}}$ are dictated by the trap electrodes, $C_{\myr{iso,k}}$, $k=1,2$ can be chosen independently.  There is an inherent trade off in this choice, however. On the one hand these should be much larger than $C_{\myr{rf,k}}$ in order to maximize the trap drive voltage. On the other hand these should be as small as possible so as to minimize $C_{\myr{trap}}$ and increase the coupling rate $g$. For simplicity, here we choose $C_{\myr{iso},1}=C_{\myr{iso},2}=C_{\myr{rf},1}=C_{\myr{rf},2}$ but other choices could be explored. For the trap in ~\Fref{fig:trapAndDet}\capfont{a}, $C_{\myr{total}}\sim 26~\un{fF}$, so $R_{\myr{det}}/L_{\myr{ion}}\sim 2\pi\times 0.7\ $MHz. See caption of~\Fref{fig:trapAndDet} for the capacitance values for both traps. The relatively large difference between the signal bandwidths calculated above and the typical signal bandwidth in a Penning trap experiment follows from the small dimensions and small capacitance of the designs considered here. 

The width of the electron signal should be compared to the frequency spread resulting from the trap anharmonicity. Using first order perturbation theory we can estimate the effects of the $r^4,r^2z^2,z^4$ terms in the trap potential (see for example~\cite{2008EPJD...50...97B}) resulting in $\le 0.5~\un{MHz}$ dispersion in the signal for both traps in \Fref{fig:electron3dtraps}, assuming the electron thermal motion equilibrates to a $4~\un{K}$ bath. 
This should contribute very little to the broadening of a single electron signal thereby simplifying its detection without the need for a more elaborate electrode design. Notice also, that the $\le 0.5~\un{MHz}$ dispersion falls within the bandwidth of the detection circuit described above, rendering the cooling induced by coupling to the detection circuit to be effective for electrons with temperatures $\le 4~\un{K}$ (energies $\le 0.34~\un{meV}$). Even in the presence of non-linearities a single electron could be detected by parametrically driving its motion and coherently detecting the resulting image currents in the detection circuit~\cite{Wineland:1973gfa}. 

The bandwidths calculated above fall in the $0.1-1~\un{MHz}$ and therefore correspond to a single electron detection time of $1-10~\mu\un{s}$. By integrating the thermal power spectral density at $R_{\myr{det}}$ over a bandwidth of $B\equiv R_{\myr{det}}/(2\pi L_e)$ centered at $\omega_0$, the total detected power will vary from $P_\myr{det}\sim 4k_b TB$ when no electron is trapped to $P_{\myr{det}}\sim 0$ when an electron is trapped~\cite{Wineland:1975kk}. This is a result of the fact that on-resonance, the electron equivalent circuit is effectively a short which shunts $R_\myr{det}$, as seen in~\Fref{fig:electronDetection}. To avoid a large noise background, an amplifier with an effective noise temperature that is $\le T$ is required. As an example, for the $T=4~\un{K}$ experiments explored here, using an amplifier with a noise temperature of $2~\un{K}$ at $\omega_0/2\pi\sim 1~\un{GHz}$ such as in~\cite{Weinreb:2007ko} could suffice, giving an estimated signal to noise ratio of one or larger in determining the variation in $P_\myr{det}$ before and after trapping.

\subsection{Parametric cooling}\label{sec:parameteric_coupling}
The low-energy electron source described in~\Sref{sec:lowenergyelectronsource} relies on the ability to cool the motion in all three spatial axes. As described there, adequate $z$-motion cooling can be achieved when the detection circuit is resonant with the $z$-motion. By parametrically coupling the radial $x$ and $y$-motion to the $z$-motion, cooling on all axes can be achieved~\cite{Gorman:2014ba}. Such a scheme has the benefit of not needing an extra radial electrode for damping or additional resonant circuitry on the existing ring electrode.

The coupling scheme in~\cite{Gorman:2014ba} was based on $xy$ and $xz$ terms in the pseudo-potential which were proportional to a voltage $U$. Time-modulating $U(t)=U_0\cos(\Delta\omega t)$ at the difference frequency $\Delta\omega=\omega_i-\omega_j$, causes energy exchange between the motion in the $i$ and $j$ axes. The traps considered in~\Fref{fig:electron3dtraps}, however, are axially symmetric and therefore should have negligibly small cross terms of that type.  We could also consider this approach by modifying the electrodes to be able to induce couplings of this form. Alternatively, a variation on this coupling scheme could be used, that incorporates the symmetry of the simpler electrode structures. To see this, we approximate the trap pseudo potential around its minimum,
\begin{eqnarray}
\phi_{\mathrm{pseudo}}&=&\frac{1}{2}m_e\left(\omega_x^2 x^2+\omega_y^2 y^2+\omega_z^2 z^2\right)\\ \nonumber
						&+&\beta x^2z^2+\gamma y^2 z^2,
\end{eqnarray}
where the $x^2y^2$ an-harmonic term is also negligible for the axially symmetric traps considered and $\beta\approx \gamma$. In terms of the harmonic ladder operators, the $x^2z^2$ cross term, for example, contains the following summands:
\begin{equation}\label{eq:anharmonic_cross_coupling}
\hbar \xi \left(a^2b^{\dagger 2}+b^2a^{\dagger 2}\right),
\end{equation}
where $a,a^\dagger$ are the $z$-motion operators and $b,b^{\dagger}$ are the $x$-motion counterparts. Coherently driving the $z$-motion at $\omega_d=2\omega_x-\omega_z$ can be described mathematically by replacing $a\mapsto \alpha e^{-i\omega_d t}+a$. Rewriting~\Eref{eq:anharmonic_cross_coupling} and neglecting fast rotating terms introduces terms of the form
\begin{equation}
2\hbar \xi \alpha \left(ab^{\dagger 2}+b^2a^{\dagger }\right).
\end{equation}
 
As an example, consider the trap design in \Fref{fig:electron3dtraps}\capfont{a}. There, in order to achieve $x-z$ coupling, $\omega_d$ should be $\sim 2\pi\times 90~\un{MHz}$. By expressing $\beta$ in terms of the pseudo-potential parameters:
\begin{equation}
\beta=\zeta \frac{2q_e^2V_{\myr{rf}}^2}{m_e\Omega_{\myr{rf}}^2d^6},
\end{equation}
where $\zeta=0.166$ is a geometric pre-factor, we can express the $x$-$z$ coupling frequency as
\begin{equation}
2\xi\alpha=\zeta\frac{\sqrt{2\hbar} q_e^3V_{\myr{rf}}^2V_d}{m_e^{3.5}\Omega_{rf}^2\omega_x\omega_z^{2.5}d^7},
\end{equation}
where $V_d$ is the drive voltage applied to the trap endcaps. For the trap in~\Fref{fig:electron3dtraps}\capfont{a} we get a rate of $2\pi\times 0.92~\un{MHz}/\un{V}\times V_d$. Therefore, a $V_d\sim 109~\un{mV}$ drive, corresponding to $\sim 3.36\ \mu\un{m}$ of motion amplitude, would render an $x-z$ coupling rate of $2\pi\times 100~\un{kHz}$. This would enable cooling of the $x$-motion on the order of that rate. With a Q-factor of $1000$ for the detection circuit, a $109~\un{mV}$ drive at $\omega_d\sim 2\pi\times 90~\un{MHz}$ would dissipate less than $10~\un{nW}$ of power, well within the cryogenic capabilities of the refrigerator. 

\subsection{Planar arrangements}
Planar chip traps have some advantages over the three-dimensional traps analyzed above. They can be easier to fabricate, require no alignment and are more suited for scalability.  Such traps, however, have a much shallower trapping potential for the same applied voltages and frequencies, as compared to three-dimensional traps. This can be mitigated by adding a cover electrode a few millimeters away from the trap chip, and applying a negative voltage~\cite{Goldman:2010ky,Kim:2010da,Schmied:2011gm}.

\begin{figure}[!hbtp]
	\begin{center}
		\includegraphics[width=0.45\textwidth]{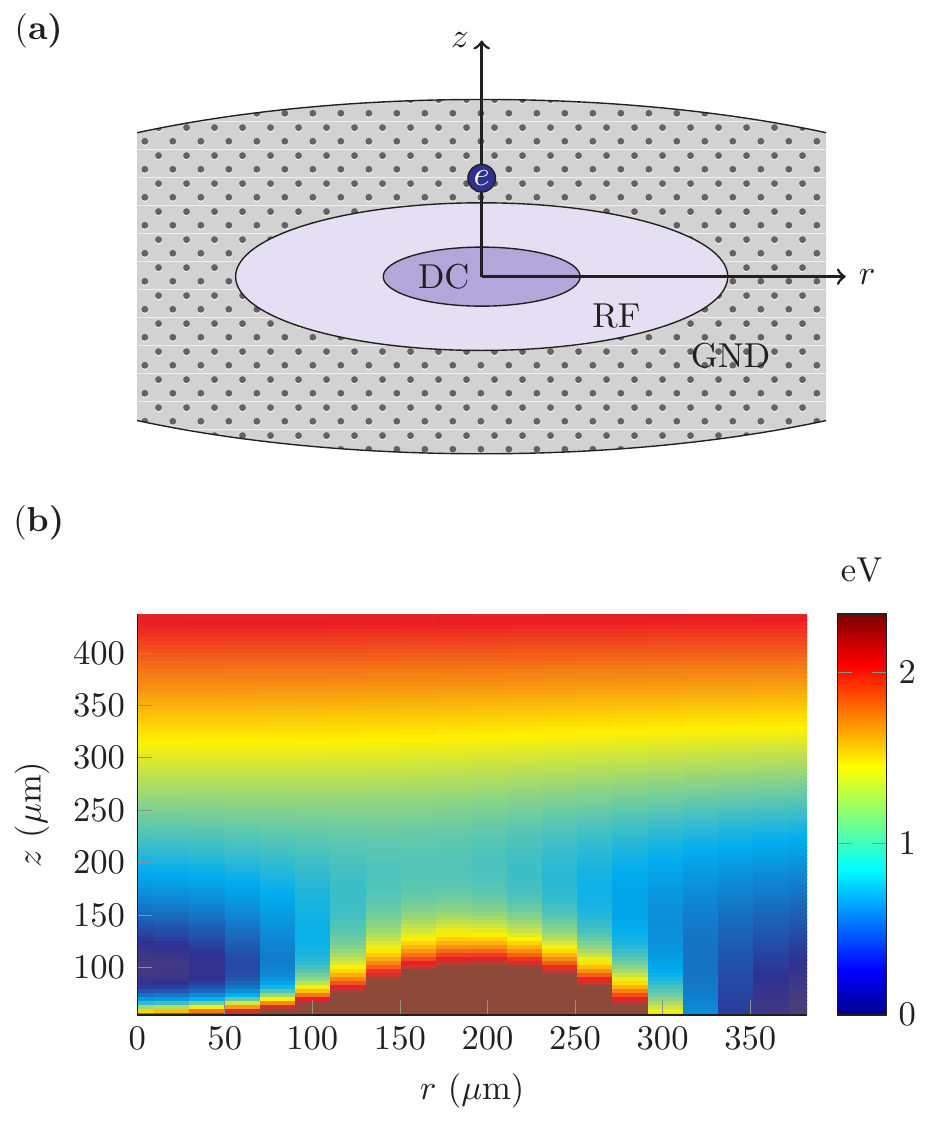}
		\caption{Planar point Paul trap for electrons. \capfont{a} Inner DC disk radius is $100\ \mu\un{m}$. Outer RF ring radius is $250\ \mu\un{m}$. The electron is trapped at a height of $\sim 100\ \mu\un{m}$ above the surface. \capfont{b} Pseudo-potential trap depth of the trap in a, with $100~\un{V}$ trap drive at $7.1~\un{GHz}$ and a capping electrode, here represented by adding a uniform field of $58.5~\un{V}/\un{cm}$ along $-z$. Trap minimum is at $r=0,\ z\sim 100~\mu\un{m}$. Resulting secular frequency along $z$ is $\omega_z=2\pi\times 1.46~\un{GHz}$.}\label{fig:planartrap}
	\end{center}
\end{figure}

Figure \ref{fig:planartrap} shows an example of a planar electrode Paul trap, here chosen to be cylindrically symmetric for simplicity. described in~\cite{Wesenberg:2008kx,Kim:2010da}. With the addition of a cover electrode generating a uniform field of $58.5~\un{V}/\un{cm}$, the trap depth is $D=0.02 q V_{\myr{rf}} q_{\myr{mathieu}}$. When applying an trap drive voltage of $V_{\myr{rf}}=100~\un{V}$ to the RF annulus electrode (DC and GND electrodes are rf-grounded) and assuming a Mathieu parameter of $q_{\myr{mathieu}}\sim 0.5$, we expect a $1~\un{eV}$ trap depth, as in the three-dimensional designs shown earlier. The relevant trap capacitance that dictates the values of the coupling rate $g$ is formed between the central dc electrode and ground. Due to the trap geometric aspect ratio, the coupling rate decreases to $g=2\pi\times 180~\un{kHz}$. 
As a side effect of using a cover electrode, the electron equilibrium position should shift towards the trap chip by $7.7\ \mu\un{m}$. This would result in $\sim 2\ \mu\un{m}$ micromotion amplitude (corresponding to a pseudo-potential energy of $\sim 15~\un{meV}$)  that should be compatible with a stable trap operation. This, however, would compromise electron loading into the trap due to the additional rf heating resulting from excess micromotion (see~\Sref{sec:lowenergyelectronsource}). One remedy could be to compensate for micromotion by applying dc voltages on the center DC electrode. In the example considered here, $1.5~\un{V}$ of dc bias would restore the ion position to the rf-null point while still rendering $1.6~\un{eV}$ deep trap.

Although planar traps seem promising, separating the detection circuit from the drive circuit  would be more difficult due to the lack of symmetry assumed in~\Sref{subsec:electricalcircuitry}. Also, since planar traps tend to be more an-harmonic compared to three-dimensional ones, additional compensation electrodes may be required in order to enable single-electron detection~\cite{Goldman:2010ky}.

\section{Concluding remarks}\label{sec:conclusion}
We have first considered coupling the motion of a confined charged particle to a superconducting resonator. Limited by the currently achieved quality factors of such resonators ($Q\le 10^6$), we conclude that for the systems considered, it will be very difficult to reach the strong coupling regime using a single trapped charged particle, with perhaps the exception of \Be at dilution-refrigerator temperatures or trapped electrons.

We explored coupling a trapped ion to a nano-mechanical resonator either through electrostatics or piezoelectricity. Based on recent advances in fabrication of membranes ($Q\ge 10^8$), we considered their electrostatic coupling to a trapped ion. By plating such a membrane with a thin metallic film and voltage biasing it, the coupling could be on the order of $10~\un{Hz}$ for a $1~\un{V}$ bias, within reach of the strong-quantum regime at $T=50~\un{mK}$.

We analyzed the possibility of direct piezo-electric coupling of ion motion to a mechanical resonator. An interesting candidate was a quartz acoustic resonator with a very high quality factor ($Q>10^9$). However, due to the relatively small overlap between the ion electric field and the acoustic mode shape, the coupling strength is found to be on the order of $1~\un{Hz}$. Reshaping the ion field with the aid of a capacitor led to an increase in the coupling, to $10~\un{Hz}$, approaching the strong quantum regime. 

By laser cooling a single \Be ion that interacts with the quartz resonator, the acoustic mode with an effective mass of $\ge 1~\un{mg}$ (!) could be cooled close to its ground state of motion. If such a massive object is placed in a superposition state, it could be used to restrict various macroscopic decoherence theories. For example, quantum gravity has been suggested to result in a motional decoherence rate that is proportional to $M^2$  for an object of mass $M$~\cite{RomeroIsart:2011es}. If a few milligram mechanical oscillator is placed in a superposition of position states differing by twice its zero-point motion, that superposition would decohere in $\sim 10~\un{ps}$. This effect should be testable since the expected coherence time of the quartz resonator is much longer, even at $4~\un{K}$. To be well within the strong quantum regime, one could engineer a different resonator, perhaps with stronger piezo-electric coefficients, that maintains a high Q factor and where the acoustic mode shape has a large overlap with the ion electric field. Such a task, however, is not straightforward as these different demands may not be compatible. 

Lastly, we considered coupling an electron to a superconducting electrical resonator. We examined two specific trap designs with a $1~\un{eV}$ trap depth, a depth we view as crucial for initial trapping where laser cooling is not available. The relatively high voltages and currents required to create such a trap depth suggest the need for thick niobium conductors to form the trap, in order to maintain superconductivity. Additionally a $1~\un{eV}$ trap requires a low-energy source of electrons, and damping to combat heating. We examined a three-dimensional trap arrangement, which can separate the high voltage, high current rf trapping circuitry from the low voltage, low currents flowing in the electron detection circuit, using trap symmetry. Obtaining a similar effect for a planar chip trap geometry would be more complicated due to the lack of symmetry.

It is worth noting the appealing properties that a hybrid system based on a trapped electron might have. Such an architecture might be more scalable compared to trapped ion QIP since the interconnecting elements are chip-based, requiring only rf control and no optical elements or laser beams. The absence of optical elements could allow for smaller traps, enabling stronger coupling between electrons and superconducting elements. Moreover, as the speed of entangling gates based on the Coulomb interaction of two charged particles scales with the trapping frequency and as a trap for electrons would typically have secular frequencies that are two orders of magnitude larger than for ions, we expect shorter electron gate times as compared to trapped ions~\cite{2008PhRvL.101i0502O}. Recent advances in entangling trapped ions have reached gate speeds which are only an order of magnitude slower than the trap frequency~\cite{2014arXiv1406.5473B,Ballance:2016hy}. If that were to scale for a trapped electron, it would correspond to a $\sim 10-100~\un{ns}$ gate time, comparable to superconducting qubit gate times~\cite{Kelly:2015gi}. Electron spin-coherence times can exceed a second~\cite{Kotler2011} and therefore be orders of magnitude larger than coherence times for superconducting qubits, where the best values to date are close to a millisecond~\cite{Reagor:2015vw}. Therefore a hybrid QIP platform based on trapped electrons might have a much larger qubit coherence time to gate time ratio. The platform might offer an additional way to entangle electrons, mediated by the underlying circuitry. This would enrich the QIP toolbox available for the electron spins. For this second method, gate speed is limited to the exchange rate between the electron and its accompanying superconducting resonator, which we estimate to be on the order of $g\sim 2\pi\times 1\ \un{MHz}$ for $50\ \mu\un{m}$ distance between electrons and superconducting circuitry and faster for smaller traps. 

\begin{acknowledgments}
	The authors would like to thank K. Cicak for her help in estimating the coupling of an ion to a membrane and for her help with electron trap design and  resulting fabrication constraints. We thank K. Bertness for discussions regarding GaN nano-beams. We thank M. Goryachev, S. Galliou and M. E. Tobar for introducing us to the physics of BVA resonators as well as lending us devices to measure.  We thank A. Sanders for introducing us to electron source and electron optics technology and his help in assessing their relevance. We thank F. Lecocq, J. D. Teufel and J. Aumentado for discussions regarding the superconducting and rf measurement aspects of this manuscript. We thank A. Sirois and D. Allcock for carefully reading this manuscript and their helpful comments.
\end{acknowledgments}

\appendix
\section{Calculating quartz resonator to ion coupling}\label{apx:quart2ion}
\begin{figure}[!hbtp]\begin{center}
		\includegraphics[scale=1]{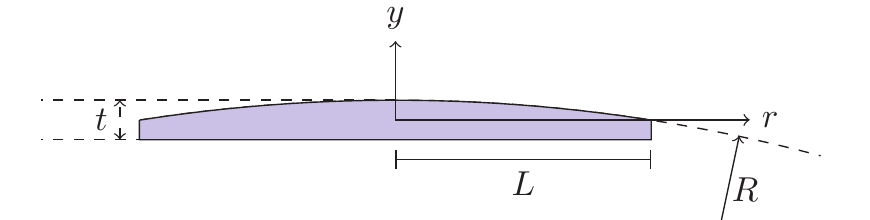}
		\caption{BVA geometry. Cylindrically symmetric about the $y$ axis with a maximal thickness $t$. The BVA lower surface is a flat disk of radius $L$. The BVA upper surface can be described by a curved surface $y= t(1-\frac{r^2}{2Rt})$  with a radius of curvature $R$. We consider a resonator (not to scale) with $R=300~\un{mm}$, $L=6.5~\un{mm}$, $t=1.08~\un{mm}$.}\label{fig:bvageom}
	\end{center}
\end{figure}

Coupling calculations require knowing the quartz resonator mode-shape $\vec{s}$, the orientation of crystallographic axes of the resonator, the corresponding $3\times 6$ piezo-electric coefficient matrix of quartz $e$, and the ion electric field. We focus on the high $Q$ modes [\Eref{eq:bvamode}] that are quasi-longitudinal, i.e. along the $\hat{n}=(0.226, 0.968, 0.111)$ unit vector, in the coordinate system described in~\Fref{fig:bvageom}. The BVA quartz resonators are made from doubly-rotated SC (stress-compensated) cut quartz~\cite{Galliou:2013br}. The coefficient matrix $e$ for this cut is taken from table 7 in the IEEE standard of piezoelectricity~\cite{Meitzler:AJ2xbzcE}.

Denote the overlap integral in the nominator of~\Eref{eq:bvamode} by $g_c$, i.e.,
\begin{equation}
g_i=\frac{\int_V d^3r \partial_i E_{ion}es'}{2\omega_0\sqrt{M m_{ion}}}\equiv\frac{g_c}{2\omega_0\sqrt{M m_{ion}}},\ i=x,y,z.\label{eq:piezocoulingagain}
\end{equation} 
The mode mass is calculated by the integral,
\begin{eqnarray}
M&=&\int_v d^3r \rho_{\myr{quartz}} \abs{s}^2 \\ \nonumber
&=& \rho_{\myr{quartz}}\pi\sigma^2\frac{t}{2}\left(1-e^{-L^2/\sigma^2}\right),
\end{eqnarray}
where $\sigma$ is the Gaussian profile radial scale of the mode shape $\vec{s}$. From~\cite{Stevens:1986im}, 
\begin{equation}
\sigma=\left(\frac{Rt^3}{3n^2\pi^2}\right)^{1/4},
\end{equation}
where $R=300~\un{mm}$ is the radius of curvature of the upper surface of the resonator, $t$ is its thickness and $n$ is the mode number (see \Fref{fig:bvageom}). An approximate formula for the resonance frequency also follows:
\begin{equation}
\omega_0=c_{\myr{sound}}\frac{n\pi}{t},
\end{equation}
where $c_{\myr{sound}}=6750~\un{m}/\un{s}$ is the speed of sound for the quasi-longitudinal modes.

An exact calculation of $g_c$ can be found in \Aref{subsec:directionbva}. Before doing so, we first estimate in~\Aref{subsec:upperbound} an upper bound on $g_c$ and correspondingly $g$, by avoiding the vector nature of the overlap integrand.

\subsection{Upper bound on direct ion-quartz coupling}\label{subsec:upperbound}
An upper bound can be obtained by using the Cauchy Schwartz inequality, applied to $g_c$:
\begin{eqnarray} 
g_c&=&\int d^3r \partial_i E_{ion} e u'\\ \nonumber
&\le& \sqrt{\int d^3r \left(\partial_i E_{ion}\right)^2\int d^3r (eu')^2}\\ \nonumber
&\le& \sqrt{\int d^3r \left(\partial_i E_{ion}\right)^2\times e_{\myr{max}}\times \int d^3r (u')^2},
\end{eqnarray}
where $e_{\myr{max}}\approx 0.234~\un{C}\cdot \un{m}^{-2}$ is the square root of the maximal eigenvalue of $e^\dagger e$. The electric field of an ion hovering at a height $h$ along the $\hat{y}$ axis is $E_{ion}(\vec{r})\approx q\vec{R}/4\pi\overline{\epsilon}R^3$ where $\vec{R}=\vec{r}-h\hat{y}$ and $\overline{\epsilon}$ is the average dielectric constant of vacuum and quartz. We can therefore write,
\begin{equation}
g_c\le \gamma  \frac{e_{max}q}{4\pi\overline{\epsilon}\sqrt{h^3}}\sqrt{\int d^3r (u')^2},\\
\end{equation}
where $\gamma$ is a numerical factor of order unity for all $i=x,y,z$ directions. 

To estimate the last integral of the strain $(u')^2$, recall that the mode mass $M=\int d^3 r \rho_{\myr{quartz}} u^2$, where $\rho_{\myr{quartz}}=2.6\times 10^3~\un{kg}/\un{m}^3$ is the quartz density. Due to the mode shape [\Eref{eq:bvamode}] we may approximate $u'\sim k u$, where $k$ is the wavenumber of the longitudinal oscillations within the BVA, i.e. $kt=n\pi$ for $t$ the resonator thickness and $n=1,3,5,\ldots$. Therefore, $\int d^3r (u')^2\sim k^2\int d^3r u^2$ and we may estimate an upper bound, 

\begin{equation}g\equiv\frac{g_c}{2\omega_0\sqrt{Mm_{ion}}}\le 
\gamma \frac{e_{\myr{max}}q}{4\pi\overline{\epsilon}c_s\sqrt{m_{\myr{ion}}\rho_{\myr{quartz}}h_0^3}}\sim 2\pi\times 1~\un{kHz}\end{equation}
where $c_s=6757~\un{m}/\un{s}$ is the speed of sound for the quasi-longitudinal mode.

\subsection{Direct ion-quartz coupling calculation}\label{subsec:directionbva}
Now that the upper bound has been established, we numerically calculate the integral in \Eref{eq:piezocoulingagain} for the low frequency modes of the quartz resonator (\Tref{tbl:ion2bvadirect}). We see that all coupling strengths are below $1.5~\un{Hz}$. 
\begin{table}[!htbp] 
		\begin{ruledtabular}
		\begin{tabular}{lllll}
			$n$ & Frequency & $g_y$                     & $g_x$                     & $g_z$                     \\ \hline
			$3$ & 9.4 MHz   & $2\pi\times 1.46~\un{Hz}$ & $2\pi\times 1.09~\un{Hz}$ & $2\pi\times 0.49~\un{Hz}$ \\
			$5$ & 15.6 MHz  & $2\pi\times 1.39~\un{Hz}$ & $2\pi\times 1.02~\un{Hz}$ & $2\pi\times 0.47~\un{Hz}$ \\
			$7$ & 21.9 MHz  & $2\pi\times 1.33~\un{Hz}$ & $2\pi\times 0.97~\un{Hz}$ & $2\pi\times 0.44~\un{Hz}$ \\
			$9$ & 28.1 MHz  & $2\pi\times 1.28~\un{Hz}$ & $2\pi\times 0.94~\un{Hz}$ & $2\pi\times 0.43~\un{Hz}$ \\ 
		\end{tabular}
		\end{ruledtabular}
		\caption{Direct coupling of a \Be ion to a BVA quartz resonator. The ion is assumed to be trapped $50\ \mu\un{m}$ above the quartz. The quartz thickness is assumed to be $1.08~\un{mm}$. Coupling strength $g_i$ for $i=x,y,z$ is the coupling strength for an ion motion along the $i$ axis. The longitudinal mode number is $n$.\label{tbl:ion2bvadirect}}

\end{table}

It is interesting to notice the weak dependence of the coupling strengths on the mode number $n$. Due to the frequency and mode mass scaling, the denominator of \Eref{eq:piezocoulingagain} scales like $\sqrt{n}$. On the other hand, because the derivative of the ion field is equivalent to a dipole field, the integrand of $g_c$ scales as $1/r^3$ whereas its Jacobian scales as $rdr$ so overall we should expect a $1/r\sim 1/\sigma\sim \sqrt{n}$ dependence, which nearly cancels the similar dependence in the denominator for the expression in $g$. Although, for very high frequency modes, $g$ should deteriorate due to high spatial frequency averaging of the ion field.

\subsection{Ion-quartz coupling via a shunt capacitor}\label{subsec:ionquartzcapacitive}
In the paper body, we estimated the coupling of the ion to the quartz resonator via a shunt capacitor, using a BVD equivalent electrical circuit. The main advantage of that approach, other than its simplicity, is that the effective capacitance of a BVA quartz resonator is a rather easily measurable quantity~\cite{Goryachev:2011tb}.

\begin{figure}[!h]\begin{center}
			\includegraphics[scale=1]{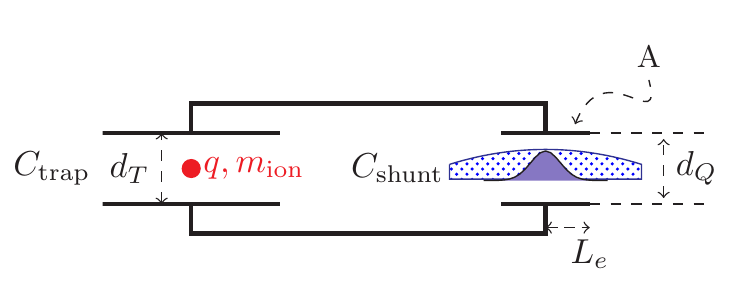}
			\caption{Coupling an ion to a quartz resonator mediated by a shunt capacitor. An ion is elastically trapped (trap electrodes no shown) at the center of a parallel plate capacitor. The ion motion generates image currents that in turn generate an electric field between the parallel plate capacitor (each plate with area $A=\pi L_e^2$) encapsulating the quartz resonator.}\label{fig:iontocapwithgeom}
		\end{center}
	\end{figure}

Here, we use \Eref{eq:piezocoulingagain} to directly calculate the coupling strength, in order to infer its dependence on mode parameters. To this end, we have to introduce parameters that describe the geometry involved. The ion is assumed to be trapped at the center of a parallel plate capacitor whose plates are a distance $d_T$ from one another (see~\Fref{fig:iontocapwithgeom}). The quartz resonator is assumed to be enclosed in another parallel plate capacitor, with a distance $d_Q$ between the plates and a plate area of $A$.

If the ion is displaced by $\Delta y$ from equilibrium towards one of the plates, it will generate an image charge $q^*=\Delta y q/d_T$.  A portion of these image charges spread uniformly on the BVA shunt capacitor plates, creating a charge density $\sigma=q^*/A(1+C_{\myr{trap}}/C_{\myr{shunt}})$ and exerting a field inside the BVA volume $E=\sigma/\epsilon$. We get:
\begin{equation}
\frac{dE}{d\Delta y}=\frac{q}{\epsilon Ad_T(1+C_{\myr{trap}}/C_{\myr{shunt}})},
\end{equation} 
where $C_{\myr{trap}}$ is the trap capacitance, and $C_{\myr{shunt}}$ is the BVA shunt capacitance and the field is perpendicular to the plates. As before, we focus on the quasi-longitudinal mode shapes [\Eref{eq:bvamode}]. Performing the overlap integral in this case results in
%
\begin{equation} 
g_c=\frac{4q\bar{e}}{\epsilon d_T}\frac{\sigma^2}{L_e^2}(1-e^{-L_e^2/2\sigma^2})\frac{1}{1+\frac{C_{\myr{trap}}}{C_{\myr{shunt}}}},
\end{equation}
where $L_e$ is the electrode radius, $\bar{e}$ is the mode-shape weighted average of $e_{22},e_{2,4},e_{26}$, i.e $\bar{e}=n_ye_{22}+n_ze_{24}+n_xe_{26}=7.43\times 10^{-2}~\un{C}\un{m}^{-2}$ and $\hat{n}=(n_x,n_y,n_z)=(-0.23,-0.97,0.1)$ is the quasi-longitudinal mode direction vector. By maximizing $g_c$ as a function of $L_e$ and for $C_{\myr{trap}}=50\ $fF trap capacitance, we estimate $L_e=1.05\sigma$ so the coupling rate is 
\begin{equation} 
g=\frac{0.58 q \overline{e}}{\epsilon d_T\omega_0\sqrt{Mm_{ion}}}=2\pi\times 10~\un{Hz},
\end{equation}
where we assumed coupling to a \Be ion, trapped between capacitor plates a distance $d_T=200\ \mu$m away from one another.

To see the geometric scaling of this, recall that $\sigma=(\frac{t^3R}{3\pi^2n^2})^{\frac{1}{4}}$ and $\omega_0\approx c_s n\pi/t$ \cite{Stevens:1986im}. We get,

\begin{equation}
g\approx 0.3\frac{q\overline{e}}{\epsilon c_s\sqrt{m_{ion}\rho_{\myr{quartz}}}}\frac{1}{d_T(tR/2)^{\frac{1}{4}}\sqrt{n}}\label{eq:geomscalingforcapBVA}.
\end{equation}

From \Eref{eq:geomscalingforcapBVA}, we expect the coupling to diminish for higher modes (increasing $n$). The dependence in the geometrical parameters $t,R$ is also very weak ($1/4$ exponent) with values limited to thicknesses in the range of $0.5-1~\un{mm}$ and radii of curvature in the $R\sim 300~\un{mm}$ range.

\section{Estimating electron ``anomalous" motional heating rate from ambient noise}\label{apx:electronmotionheating}
We estimate the ``anomalous" heating rate of the electron motion by extrapolating from known ion heating rates~\cite{Turchette2000,Hite:2013iw,Brownnutt:2015io}. If $n$ denotes the average number of motional quanta in a trap with frequency $f$ then
\begin{equation}
\dot{n}\propto\frac{q^2}{m}\frac{1}{d^4f^{1+\alpha}}\label{eq:heatingrate},
\end{equation}
where $q,m$ are the particle charge and mass respectively and $\alpha$ has varied between $0.5$ and $2$ in various experiments. In this expression, $d$ is the distance of the charge from the nearest electrode and we assume the electric field noise is generated by independent fluctuating patch potentials of extent $< d$~\cite{Turchette2000}.

From \Tref{tbl:heatingrates} and \Eref{eq:heatingrate} we can estimate the electron heating rate to be between $30-160~\un{quanta}/\un{s}$ for a trap-to-electron distance of $\sim 50\ \mu\un{m}$ and an electron motional frequency of $\sim 1~\un{GHz}$, assuming $\alpha=0.5$. For $\alpha=2$, all of the extrapolated heating rates are below $0.02~\un{quanta}/\un{s}$. 
These rates are at least three orders of magnitude smaller than the coupling rates we expect between the electron and the superconducting resonator. Specifically, with the traps considered in this paper the coupling rates were estimated to be in the range $g/2\pi=180~\un{kHz}-1.06~\un{MHz}$.

     \begingroup
     \squeezetable
\begin{table} 
		\begin{ruledtabular}
		\begin{tabular}{lllllll}
			Trap material                 & T            & ion           & $d$              & $f$             & $\dot{n}$               & Ref. \\ 
			\hline
			Au on sapphire              & $5~\un{K}$   & $^{88}$Sr$^+$ & $50\ \mu\un{m}$  & $1.32~\un{MHz}$ & $4~\un{quanta}/\un{s}$  & \cite{Chiaverini2014}\\
			Au on quartz & $300~\un{K}$ & \Be           & $40\ \mu\un{m}$  & $3.6~\un{MHz}$  & $58~\un{quanta}/\un{s}$ & \cite{Hite:2012cu}\\
			Nb on sapphire           & $6~\un{K}$   & $^{88}$Sr$^+$ & $100\ \mu\un{m}$ & $1~\un{MHz}$    & $2~\un{quanta}/\un{s}$  & \cite{Wang:2010fa}
		\end{tabular}
		\end{ruledtabular}
		\caption{Selected measured heating rates $\dot{n}$ for ion traps. Ion to surface distance is $d$, $f$ is the trap frequency.\label{tbl:heatingrates}}
\end{table}
\endgroup

\section{Electron heating rate due to incoming electrons during the loading process}\label{apx:eecrosssection}
We estimate an upper bound for the heating rate of trapped electrons due to collisions with incoming electrons during trap loading. We assume that a single electron is trapped in a three-dimensional harmonic potential with $\sim 1~\un{GHz}$ secular frequency in all axes with a trap depth of $U_{\myr{depth}}=1~\un{eV}$. Incoming electrons, each having $E_{p}=30~\un{eV}$ of kinetic energy, collide with the trapped electron causing heating. 

We focus on a single trapped electron collision process, since we are aiming at a steady state number of just one to few trapped electrons. Moreover, we assume that the trapped electron interacts with just one incoming electron at a time. This is consistent with the incoming-electron current values we considered in \Sref{sec:lowenergyelectronsource} and the time scale for the collision process (see below).

We ignore the trap dynamics during any single collision since the former is relatively slow compared to the latter. To see this, first note that the time scale for a collision process is $b/v_{p,0}$ where $b$ is the impact parameter and $v_{p,0}$ is the incoming electron initial velocity. The impact parameter is limited by the overall incoming electron beam radius $r_0$, which we assume is $<100\ \mu\un{m}$. The incident electron speed is $v_{p,0}=\sqrt{2E_p/m_e}=3.2\times 10^6~\un{m}/\un{s}$ where $m_e$ is the electron mass. Therefore the collision duration times are $\le 3\times 10^{-11}~\un{s}$, i.e. shorter than the trap drive period ($\sim 10^{-10}~\un{s}$) and much shorter than the trap harmonic period ($\sim 10^{-9}~\un{s}$). Based on our trap parameters, we can estimate that during a collision, trap forces will change the positions of the two electrons by no more than $\sim 20\ \%$ as compared to a collision where no trap is involved. Since we are interested only in an order-of-magnitude estimate, we ignore these deviations from a trap-free calculation. 

For our purposes, however, the trap still plays a role in determining the initial conditions of the collision process. Trapped electrons have an initial energy below $U_{\myr{depth}}$. For simplicity we assume that the initial energy distribution is uniform in the range $0\le E_{s,0}\le U_{\myr{depth}}$ (see for example figure 5 in~\cite{Grissom:1972bm}). The incoming electron, at the moment of entrance into the trapping region, either accelerates or decelerates prior to the collision, depending on the phase of the trap drive. For concreteness we use the geometry in~\Fref{fig:electron3dtraps}\capfont{b}, trap parameters of~\Tref{tbl:electron3dtraps}, and assume that the incoming electrons velocity is initially along the trap symmetry axis ($z$). The incoming electron's initial kinetic energy prior to collision will be spread by $\pm 15~\un{eV}$ around $E_p=30 ~\un{eV}$, as we show later. Since the primary electron beam is initially aligned parallel to the rf electric field, the rf-trap-induced spread in $E_p$ is maximal. If, for example, the electrons come at an angle of $\sim 54.7^\circ$ with respect to $z$, the energy spread in $E_p$ reduces to $\pm 2.5~\un{eV}$. At this angle, to first order, the rf-trap field lines are perpendicular to the incoming electrons initial velocity. Our choice of geometry and electron direction therefore accentuates the spread in $E_p$ due to the rf in order to fully appreciate its influence on the heating rate. Another effect of the trap is electron deflection in the transverse direction resulting in a rastering of the incoming beam. It can be shown using elementary electrostatic consideration that the beam radius will expand by $\le \exp(2\arcsin{(q_eV_{\un{rf}}/(E_p+eV_{\un{rf}}))})< 4$. Therefore, one must make sure that the initial beam diameter is small enough such that the beam does not strike the trap electrodes from rastering.

We assume that the process can be reasonably captured by classical mechanics. We therefore ignore the spins of the electrons, as well as, scattering interference effects. The ratio between the quantum mechanical differential cross section for electron-electron Coulomb scattering $\left(d\sigma/d\Omega\right)_{\myr{quantum}}$ and its classical counterpart $\left(d\sigma/d\Omega\right)_{\myr{classical}}$ can be bounded by $0.5<\abs{\left(d\sigma/d\Omega\right)_{\myr{quantum}}/\left(d\sigma/d\Omega\right)_{\myr{classical}}}<1.03$, based on our parameters~\cite{*[{see e.g. }] [{ p. 300, Eq. (2), as well as }] Mott:104647,*Mott:1930jl}. The quantity of interest is the energy gain per collision, $\Delta E$, which is the average of the energy gained per scattering direction over an appropriate range of solid angle. Therefore, our classical estimation of $\Delta E$ will also not deviate from a full quantum-mechanical estimation by more than the above bounds.
	

\begin{figure}[!hbtp]
	\begin{center}
		\includegraphics{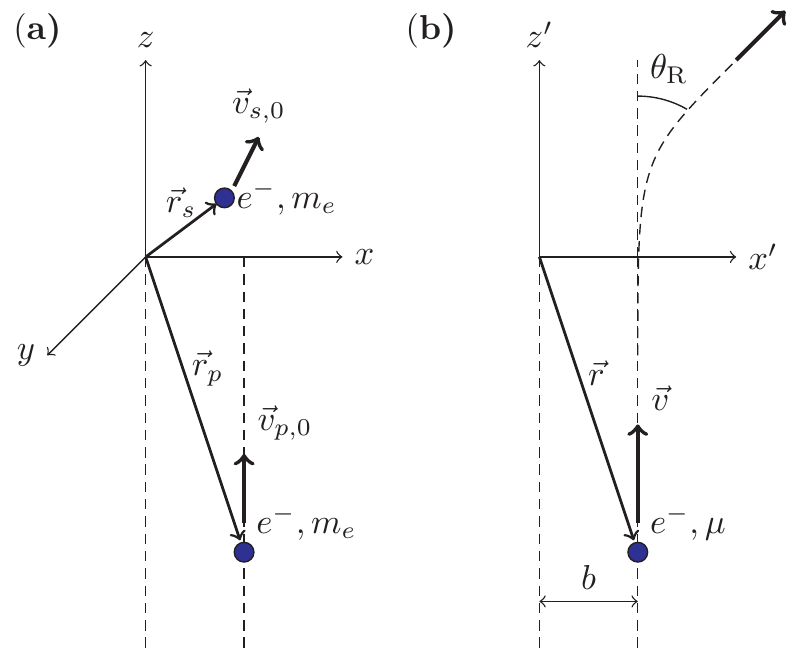}
	\end{center}
	\caption{Geometry of electron-electron scattering. \capfont{a} Lab frame. An incoming fast electron with velocity $\vec{v}_{p,0}$ collides with a slow (trapped) electron with velocity $\vec{v}_{s,0}$. \capfont{b} Reduced mass frame of reference. Here, $\vec{r}\equiv \vec{r}_{p}-\vec{r}_{s}$,  $\vec{v}\equiv  \vec{v}_{p,0}-\vec{v}_{s,0}$ and $\mu=m_e/2$ is the reduced mass. The angle $\theta_{\mathrm{R}}$ is the deflection angle of $\vec{v}$ with respect to its initial direction, after the collision.}\label{fig:eescattering}
\end{figure}

The geometry of a collision process is shown in \Fref{fig:eescattering}\capfont{a}. An incoming electron with velocity $\vec{v}_{p,0}$ and position $\vec{r}_p$ collides with a relatively slow trapped electron (the target electron) with velocity $\vec{v}_{s,0}$ and position $\vec{r}_s$. Our subscripts follow the convention of electron scattering terminology where the incoming electrons are called ``primary'' whereas the (possibly) scattered electrons are called ``scattered''. The scattering problem can be described in the center of mass and reduced mass coordinates: $R_{\myr{cm}}\equiv (\vec{r}_p+\vec{r}_s)/2$, and $\vec{r}\equiv \vec{r}_p-\vec{r}_s$, respectively. Ignoring the trapping potential as mentioned above, we can assume that the center of mass will move at a constant velocity of $\vec{V}_{\myr{cm}}=(\vec{v}_{p,0}+\vec{v}_{s,0})/2$. The relative motion is described in the primed coordinate system shown in \Fref{fig:eescattering}\capfont{b}. It is subsequently reduced to a Rutherford scattering problem of a particle of one electron charge and a reduced mass of $\mu=m_e/2$, moving with an initial velocity $\vec{v}=\vec{v}_{p,0}-\vec{v}_{s,0}$ and an impact parameter $b$, in the Coulomb potential of a fixed electron at the origin [see~\Fref{fig:eescattering}\capfont{b}]. The relative velocity vector will therefore be deflected with respect to its initial direction by
\begin{equation}
\theta_{\myr{R}}=2\arctan\left(\frac{q_e^2/4\pi\epsilon_0 b}{\mu v^2} \right),
\end{equation}
where $v\equiv |\vec{v}|$. 

Returning to the lab frame, the target electron final velocity is
\begin{equation}\label{eq:vsfinal}
\vec{v_{s}}=\frac{\vec{v}_{p,0}+\vec{v}_{s,0}}{2}-\frac{\vec{v}\cos\theta_{\myr{R}}+v\hat{u}\sin\theta_{\myr{R}}}{2},\\
\end{equation}
where
\begin{equation}
\hat{u}=\frac{\vec{r}-(\vec{r}\cdot \hat{v})\hat{v}}{|\vec{r}-(\vec{r}\cdot \hat{v})\hat{v}|},\quad\hat{v}=\frac{\vec{v}}{v}.
\end{equation}

Using \Eref{eq:vsfinal} and the triangle inequality we can find an upper bound for $|\vec{v}_s|$:
\begin{eqnarray}
|\vec{v}_s|&\le& |\vec{v}_{p,0}-\vec{v}_{s,0}|\left|\sin\frac{\theta_{\myr{R}}}{2}\right|+|\vec{v}_{s,0}|\\ \nonumber
&\le& \left(1+\sqrt{\frac{E_{\myr{thresh}}}{E_p}}\right)\left|\vec{v}_{p,0}\right|\left|\sin\frac{\theta_{\myr{R}}}{2}\right|+|\vec{v}_{s,0}|,
\end{eqnarray}
where $E_{\myr{thresh}}$ is the maximal energy of an initially trapped electron (see~\Sref{sec:lowenergyelectronsource}). This translates into a bound on the change in the kinetic energy of the target electron:
\begin{eqnarray}
\abs{\Delta E}&=&\abs{\frac{1}{2}m_e |\vec{v}_{s}|^2-\frac{1}{2}m_e |\vec{v}_{s,0}|^2}\\ \nonumber
&\le& \gamma E_p\left|\sin\frac{\theta_{\myr{R}}}{2}\right|,
\end{eqnarray}
where
\begin{equation}
\gamma=\left(1+\sqrt{\frac{E_{\myr{thresh}}}{E_p}}\right)\left(1+3\sqrt{\frac{E_{\myr{thresh}}}{E_p}}\right) \label{eq:eegamma}.
\end{equation}
If we use $\Ud$ as a bound for $E_{\myr{thresh}}$, we get $\gamma\approx 1.83$. However, in~\Sref{sec:lowenergyelectronsource} we showed that only electrons with $E_{\myr{thresh}}=0.3~\un{meV}$ are expected to be trapped, corresponding to $\gamma\approx 1.01$. 
The average change in the absolute value of the target electron kinetic energy is therefore,
\begin{equation}\label{eq:energyperkick}
\langle\abs{\Delta E}\rangle\le \gamma\frac{q_e^2}{4\pi\epsilon_0r_0},\\
\end{equation}
where $r_0$ is the incoming electron beam radius. Here, we averaged over all possible impact parameters $b$, assuming that the incoming electrons are uniformly distributed in an electron beam having a radius of $r_0$:
\begin{eqnarray}\label{eq:approxSinThetaR}
\nonumber
\left\langle \gamma E_p\left|\sin\frac{\theta_{\myr{R}}}{2}\right| \right\rangle &=& \frac{\gamma E_p}{2r_0^2}\int_0^{2r_0}db b \frac{1}{\sqrt{1+(\frac{2\pi\epsilon_0bm_e v^2}{q_e^2})^2}}\\ \nonumber
&\approx& \frac{\gamma E_p}{2r_0^2}\int_0^{2r_0}db b \frac{1}{\sqrt{1+(\frac{4\pi\epsilon_0bE_p}{q_e^2})^2}}\\
&\approx&\gamma \frac{q_e^2}{4\pi\epsilon_0 r_0}
\end{eqnarray}
where the approximation $v\sim v_p$ was used. 

\begin{figure}[!hbtp]
	\begin{center}
		\includegraphics[scale=1]{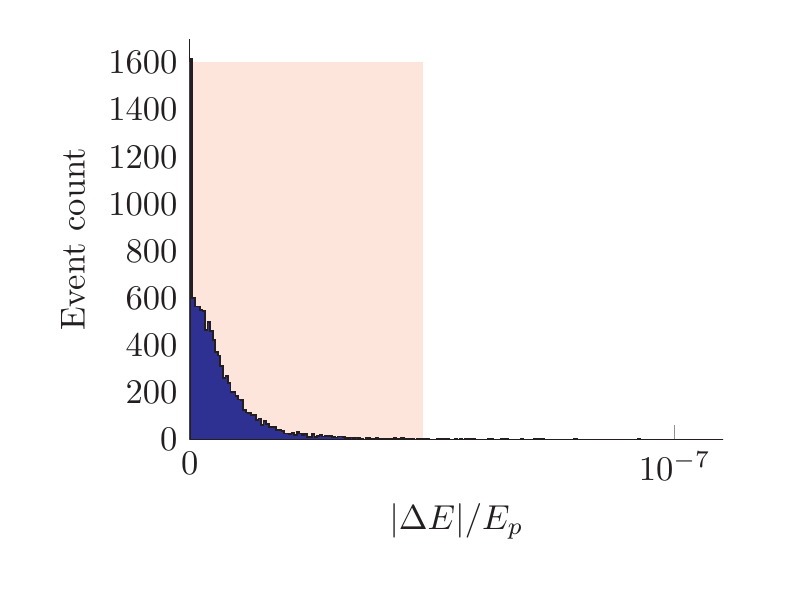}
	\end{center}
	\caption{Histogram of the absolute value of the change in total energy of a trapped electron, $\abs{\Delta E}$, due to collisions with an $E_p=30~\un{eV}$ incoming electrons. Shaded pink region shows the analytic bound in \Eref{eq:energyperkick}. We numerically integrate the equations of motion for an electron, trapped initially at $x=y=z=0$, with initial energy $E_s$ interacting with the incoming electron. We assume a pseudo-potential harmonic trap with $1~\un{GHz}$ frequencies in all axes in a trapping volume of $l^3=(95\ \mu\un{m})^3$. Here we do not account for micro-motion dynamics. The energy $E_s$ is assumed to be uniformly distributed between $0\le E_s\le U_{\myr{depth}}=1.01~\un{eV}$ and the incoming electron position is assumed to be at $z=-1~\un{mm}$ with $x,y$ uniformly distributed in the beam cross-section, $\sqrt{x^2+y^2}\le r_0=100\ \mu\un{m}$. }\label{fig:validateee}
\end{figure}

A subtle point in the calculation of the average in \Eref{eq:approxSinThetaR} is the assumption of a uniformly distributed (spatial) incident electron beam. While this assumption is reasonable in the lab frame, it is not immediately clear that it is adequate for the center of mass frame. For trapped electrons with an initial energy $\le E_{\myr{thresh}}=0.3~\un{meV}$, that is, significantly smaller than $E_p=30~\un{eV}$, the assumption of uniformity is a good approximation since the lab frame and center of mass frame are nearly identical. The value of $E_{\myr{thresh}}$ might be larger if measures are taken to decrease trap anharmonicity. The ultimate bound for $E_{\myr{thresh}}$ is therefore $\Ud$. In that case, we can see numerically that going to the center of mass frame redistributes the impact parameters to include a larger range of distances and consequentially a lower average impact energy. The calculation in \Eref{eq:approxSinThetaR} can therefore be regarded as an upper bound on the actual average value of $|\sin(\theta_{\myr{R}}/2)|$. As an example, we compare this bound to a histogram of $\abs{\Delta E}$ derived from a numerical integration of the collision equation of motion for a random set of initial conditions, as seen in \Fref{fig:validateee}. The target electron energy before collision $E_{s,0}$ is assumed to be uniformly distributed $0\le E_{s,0}\le U_{\myr{depth}}$. The incoming electron beam is assumed to be uniformly distributed. From the histogram, the average absolute value of the energy imparted to the target electron per collision is $\sim 0.74\times 10^{-7} E_p$. Assuming $0\le E_{s,0}\le U_{\myr{thresh}}$, this average decreases to $\sim 10^{-9}E_p$. Both values are consistent with the analytic expression in \Eref{eq:energyperkick} which yields a bound of $4.8\times 10^{-7}E_p$. The simulation is set up to account for the effect of the trapping pseudo-potential during the collision process, thereby serving as an independent validation of the omission of trap dynamics in our analytic derivation. 

Since the bound in \Eref{eq:energyperkick} does not depend on the target electron initial velocity, it can be translated to a corresponding average heating rate bound by multiplying it by the incoming rate of electrons. A current density of $J$ incoming electrons results in $J\pi r_0^2/q_e$ collisions per second, which in turn results in a heating rate bound of:
\begin{equation}\label{eq:dEdtduetoee}
\left(\frac{dE}{dt}\right)_e \le \frac{q_eJr_0}{4\epsilon_0},
\end{equation}
where we approximated $\gamma\approx 1$.  

The above discussion did not include the effect of micromotion on the collisions. The effect of the trap drive is to spread the kinetic energy of the incoming electron as well as the impact parameter of the collision. The bound in \Eref{eq:dEdtduetoee} changes only by a factor of order unity due to micromotion. To see this, we first consider the simple case of a target electron initially at rest in the absence of rf fields. Using \Eref{eq:vsfinal} we can write the target electron exact final kinetic energy due to a single collision:
\begin{equation}\label{eq:naiveee}
E_s=E_p\frac{x^2}{1+x^2},\qquad x\equiv \frac{q_e^2/4\pi\epsilon_0 b}{E_p}.
\end{equation}
For $x\ll 1$ (equivalently $b\gg 1\ {\angstrom}$), faster (slower) incoming electrons result in a smaller (larger) increase of the target electron energy: $E_s\propto 1/E_p$.

\begin{figure*}
		\includegraphics[scale=1]{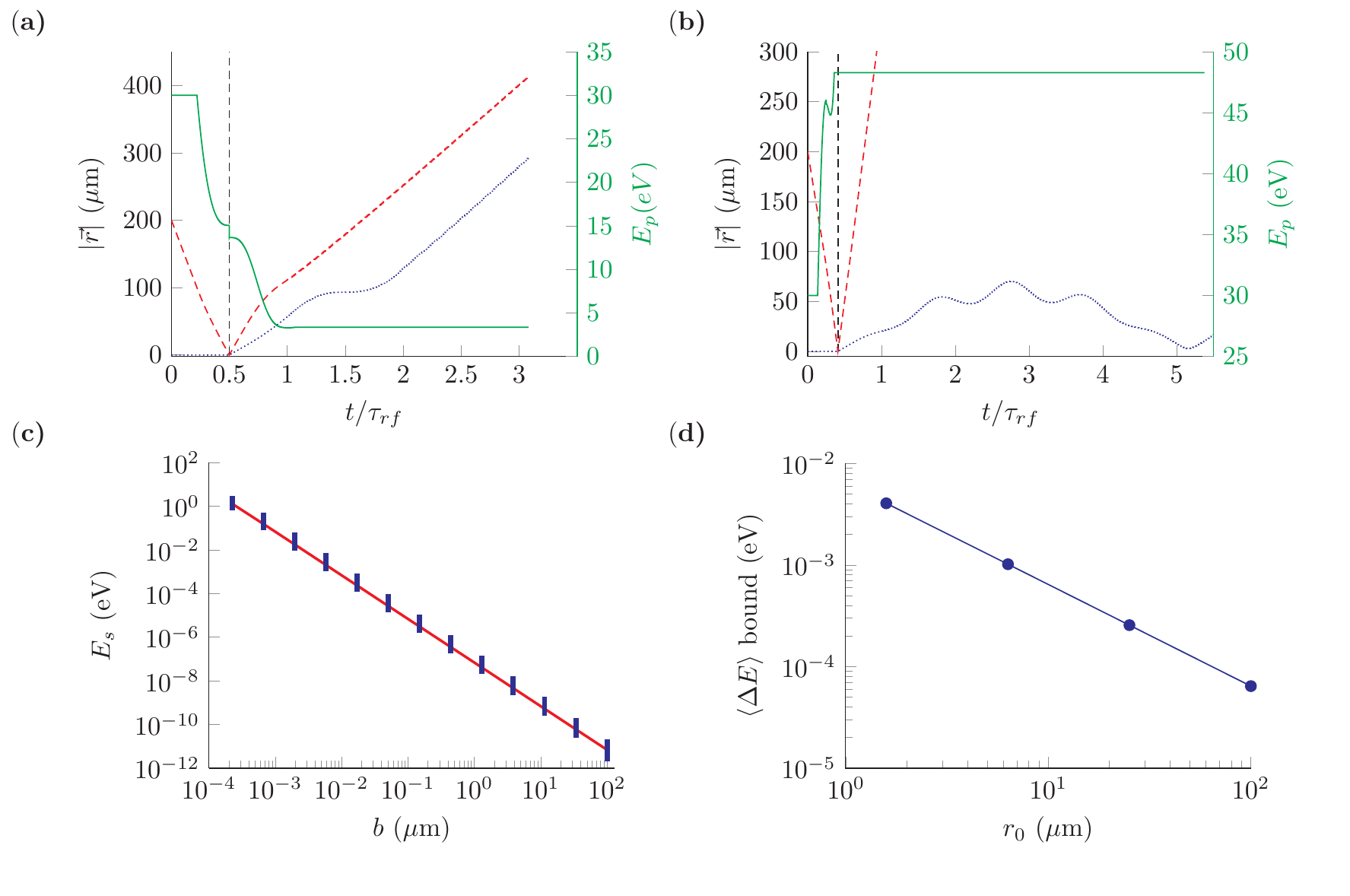}
	\caption{The effect of micromotion on electron-electron scattering for a harmonic trap with dimensions and frequencies similar to those described in \Fref{fig:electron3dtraps}b. \capfont{a} Example of a simulated trajectory of a trapped electron (dotted blue curve) colliding with an incoming electron (dashed red curve) at an impact parameter $b=10\ {\angstrom}$, as a function of time in units of the trap rf period $\tau_{rf}$. Trap center is assumed at the origin $x=y=z=0$ and $\vec{r}$ is the particle position. The instantaneous kinetic energy of the incoming electron $E_p$ (solid green curve) decreases prior to the collision due to the varying rf potential. When the incoming electron is at a distance on the order of $\sim b$ from the trap center (dashed black line), the incoming electron looses $1.39~\un{eV}$  giving the trapped electron enough energy to escape the trap. \capfont{b} Same as \capfont{a}, for an initial rf phase shifted by $\pi$ radians as compared \capfont{a}. In this case, the trapped electron gains $0.27~\un{eV}$ due to the collision, resulting in confined oscillations. \capfont{c} Blue vertical lines show the spread in the final target electron energy $E_s$ vs. impact parameter $b$, resulting from different initial trap rf phases. Analytic theory of \Eref{eq:naiveee} is shown by the solid red line. \capfont{d} Bound on the average energy gain per collision vs. incoming electron-beam radius $r_0$. Target electron initial kinetic energy is assumed to be uniformly distributed from $0~\un{eV}$ to $1~\un{eV}$. Analytic theory of \Eref{eq:approxSinThetaR} (solid blue line) is compared to a numerical integration of \Eref{eq:approxSinThetaRwithmm} that includes the spread in impact parameters and incoming electron kinetic energies due to micromotion (blue circles). The spread in these values is calculated by numerical integration of the equations of motion for the two electrons, for various initial conditions, under the influence of the trap rf field as well as their Coulomb repulsion. Initial conditions are assumed uniform as in~\Fref{fig:validateee}.}\label{fig:eewithmm}
\end{figure*}

In the presence of an rf trap, the incoming electron can either accelerate or decelerate before the collision, depending on the initial phase of the trap drive when it entered the trapping region. An accelerated (decelerated) electron will therefore transfer less (more) energy to the target electron as compared to the no-trap collision. This is exactly the case for the two examples shown in \Fref{fig:eewithmm}\capfont{a} and \capfont{b}. These simulate collision processes for initial rf phases that differ by $\pi$ radians. For concreteness, we assumed an rf trap with dimensions and frequencies as in \Fref{fig:electron3dtraps}\capfont{b} and \Tref{tbl:electron3dtraps}. We simplified the calculation by assuming the trap is harmonic in the entire cylindrical volume bounded by the electrodes. The incoming electron initial velocity is assumed parallel to the trap $z$ axis. Figure \ref{fig:eewithmm}\capfont{a} shows a collision process where the incoming electron is maximally decelerated to a kinetic energy of $E_p=15~\un{eV}$ at the beginning of the collision. This results in the ejection of the target electron from the trap. Figure \ref{fig:eewithmm}\capfont{b} shows the other extreme case where the incoming electron experiences maximal acceleration resulting in $E_p=46~\un{eV}$ so the target electron remains trapped. Although this may seem paradoxical, it follows immediately from~\Eref{eq:naiveee} for impact parameters which satisfy $b\gg 1\ {\angstrom}$.

To see how well this explanation encapsulates the effect of micromotion for the general case, we compare the theory in \Eref{eq:naiveee} to the values of $E_s$ extracted from numerical simulations as a function the impact parameter $b$. We vary the values of $b$ from $1 {\angstrom}$, below which collisions are essentially head-on [equivalently $x\sim 1$ in \Eref{eq:naiveee}], to $100\ \mu\un{m}$, i.e. the electron-beam radius. For a given value of $b$, the different values of the trap initial rf phase result in the spread in $E_s$ values shown in \Fref{fig:eewithmm}\capfont{c} (blue x markers). The center of these distributions, however, follows the theory in \Eref{eq:naiveee}, which assumes no trap drive [solid red line in \Fref{fig:eewithmm}\capfont{c}]. Overall, the effect of micromotion is a $\sim 60 \%$ spread in the value of $E_s$, centered at the value given by \Eref{eq:naiveee}.

Finally, we extend our treatment to include non-zero initial velocity for the target electron. To this end, we repeat the calculation in \Eref{eq:approxSinThetaR} with the addition of averaging over the rf initial phase, $\phi_{rf}$:
\begin{multline}\label{eq:approxSinThetaRwithmm}
\left\langle \gamma E_p\left|\sin\frac{\theta_{\myr{R}}}{2}\right| \right\rangle =\\ 
=\frac{\gamma}{2\pi r_0^2}\int_0^{r_0}db b \int_{0}^{2\pi}d\phi_{rf}\frac{E_{p,\myr{col}}}{\sqrt{1+(\frac{2\pi\epsilon_0b_{\myr{col}}m_e v^2}{q_e^2})^2}},
\end{multline}
where $E_{p,\myr{col}}$ and $b_{\myr{col}}$ are the impact energy and impact parameter at the time of collision and are functions of $\phi_{rf}$. Numerical evaluations of \Eref{eq:approxSinThetaRwithmm} are in good agreement with the analytic theory which assumed the absence of a trap [\Eref{eq:approxSinThetaR}], as can be seen in \Fref{fig:eewithmm}\capfont{d}. Here, the value of $\gamma$ in \Eref{eq:eegamma} changes to $\gamma\sim 2.23$ to account for the maximally decelerated incoming electrons, with energies as low as $15~\un{eV}$.

\end{document}